\documentclass[journal]{IEEEtran}
\usepackage{amsmath,amsfonts}
\usepackage{algorithmic}
\usepackage{algorithm}
\usepackage{array}
\usepackage[caption=false,font=footnotesize,labelfont=rm,textfont=rm]{subfig}
\usepackage{textcomp}
\usepackage{stfloats}
\usepackage{url}
\usepackage{verbatim}
\usepackage{graphicx}
\usepackage{cite}
\usepackage{xcolor}
\usepackage{bm}
\usepackage{multirow}
\usepackage{makecell}
\usepackage{xpatch}
\makeatletter
\def\changeBibColor#1{
  \in@{#1}{Liu2022Integrated,Liu2022A,Chen2021Distributed,Nuss2017MIMO,Favarelli2023Sensor,Wang2024Joint}
  \ifin@\color{black}\else\normalcolor\fi
}
 
\xpatchcmd\@bibitem
  {\item}
  {\changeBibColor{#1}\item}
  {}{\fail}
 
\xpatchcmd\@lbibitem
  {\item}
  {\changeBibColor{#2}\item}
  {}{\fail}
  
\makeatother

\begin{document}

\title{Jointly Optimizing Terahertz based Sensing and Communications in Vehicular Networks: A Dynamic Graph Neural Network Approach}


\author{Xuefei~Li,~\IEEEmembership{Student Member,~IEEE,} Mingzhe~Chen,~\IEEEmembership{Member,~IEEE,} Ye~Hu,~\IEEEmembership{Member,~IEEE,}
Zhilong~Zhang,~\IEEEmembership{Member,~IEEE,}
Danpu~Liu,~\IEEEmembership{Senior Member,~IEEE,}
and~Shiwen~Mao,~\IEEEmembership{Fellow,~IEEE}
\vspace{-0.8cm}

\thanks{X. Li, Z. Zhang, and D. Liu are with the Beijing Laboratory of Advanced Information Network, Beijing University of Posts and Telecommunications, Beijing, 100876, China (e-mail: 2013213202@bupt.edu.cn; zhilong.zhang@outlook.com; dpliu@bupt.edu.cn).}
\thanks{M. Chen is with the Department of Electrical and Computer Engineering and Institute for Data Science and Computing, University of Miami, Coral Gables, FL, 33146, USA (e-mail: mingzhe.chen@miami.edu).}
\thanks{Y. Hu is with the Department of Industrial and Systems Engineering, University of Miami, Coral Gables, FL, 33146, USA (email: yehu@miami.edu).}
\thanks{S. Mao is with the Department of Electrical and Computer Engineering, Auburn University, Auburn,
AL 36849, USA (e-mail: smao@ieee.org).}}



\maketitle

\begin{abstract}
In this paper, the problem of vehicle service mode selection (sensing, communication, or both) and vehicle connections within terahertz (THz) enabled joint sensing and communications over vehicular networks is studied. The considered network consists of several service provider vehicles (SPVs) that can provide: 1) only sensing service, 2) only communication service, and 3) both services, sensing service request vehicles, and communication service request vehicles. Based on the vehicle network topology and their service accessibility, SPVs strategically select service request vehicles to provide sensing, communication, or both services. This problem is formulated as an optimization problem, aiming to maximize the number of successfully served vehicles by jointly determining the service mode of each SPV and its associated vehicles. To solve this problem, we propose a dynamic graph neural network (GNN) model that selects appropriate graph information aggregation functions according to the vehicle network topology, thus extracting more vehicle network information compared to traditional static GNNs that use fixed aggregation functions for different vehicle network topologies. Using the extracted vehicle network information, the service mode of each SPV and its served service request vehicles will be determined. Simulation results show that the proposed dynamic GNN based method can improve the number of successfully served vehicles by up to 17\% and 28\% compared to a GNN based algorithm with a fixed neural network model and a conventional optimization algorithm without using GNNs.
 
\end{abstract}


\section{Introduction}
 \IEEEPARstart{S}{ince} the joint design of wireless sensing and communications on a single hardware platform can improve spectral efficiency and reduce hardware complexity, it is considered as a promising technology to support various vehicular applications (e.g., autonomous driving and vehicle platooning) \cite{Liu2022Integrated,Liu2022A,Tong2023Multi,Chen2021Distributed}. The shortage of wireless spectrum in sub-6 GHz band can substantially constraint the performance of the joint sensing and communication services, especially for the vehicular applications where the densely deployed moving vehicles request frequent joint sensing and communication services. This leads us to the usage of high frequency band spectrum, especially to the usage of wider, extra high data rate terahertz (THz) band.
 However, the sensing and communication signals transmitted in THz bands experiences much higher path loss and are highly vulnerable to blockages \cite{Han2022THz,Elbir2022Terahertz,Yan2023Dynamic}. Hence, using THz band for vehicular networks to provide high-resolution sensing and high data rate communication services faces many challenges such as compensation for severe path loss, reduction of link blockages, adaptation to dynamic vehicle network topology, and meeting various sensing and communication requirements.

Recently, several works in \cite{Zhang2022Time, Yao2022Intelligent,Cong2023Vehicular, Ma2023Performance,Fan2023Radar,Cheng2022Integrated,Favarelli2023Sensor}, have studied the problems of using radio frequency for both communications and sensing over vehicular networks,  as summarized in Table \uppercase\expandafter{\romannumeral1}. The author in \cite{Zhang2022Time} and \cite{Yao2022Intelligent} optimized time slot allocations for sensing and communication services. The work in \cite{Cong2023Vehicular} designed a radar-assisted beamforming scheme while considering the mobility of vehicles. The authors in \cite{Ma2023Performance} achieved high-efficient communication and obstacle detection for urban autonomous vehicles by considering the channel sparsity characteristics of the joint communication and sensing systems. In \cite{Fan2023Radar}, the authors optimized the power allocation and relay vehicle selection for sensing and communication services to minimize the total power consumption in multi-hop vehicle-to-vehicle (V2V) networks. 
The work in \cite{Cheng2022Integrated} and \cite{Favarelli2023Sensor} explored the tradeoff between sensing accuracy and communication throughput by optimizing the power and time resources for sensing and communication. However, the works in \cite{Zhang2022Time, Cong2023Vehicular,Ma2023Performance,Fan2023Radar,Yao2022Intelligent,Cheng2022Integrated,Favarelli2023Sensor} may sacrifice the performance of sensing or communication service to enhance the performance of another one, since both sensing and communication services need to use limited wireless spectrum.

 To overcome this challenge, the use of higher frequency bands (e.g., THz bands) with abundant bandwidth can be a promising solution to achieve high-quality sensing and communication services simultaneously. In \cite{Mao2022Waveform}, the authors achieved accurate target sensing and high-rate communication at the same time by designing the waveform for joint sensing and communication system in low THz bands. The work in \cite{Chaccour2022Joint} utilized the abundant bandwidth and directional transmission of THz bands to simultaneously provide a millimeter-level environmental sensing capability and extremely high data rates for virtual reality (VR) service. The authors in \cite{Chang2022Integrated} investigated an integrated scheduling method of sensing, communication, and control for THz communications in unmanned aerial vehicle (UAV) networks by considering the THz channel particularities, reduction of link blockages, and the various service requirements. In \cite{Wu2023Sensing}, the authors designed a sensing integrated discrete Fourier transform spread orthogonal frequency division multiplexing (SI-DFT-s-OFDM) waveform for THz integrated sensing and communication system to overcome the high free space path loss, reflection loss, Doppler effects, and phase noise of THz bands. The work in \cite{Gao2023Integrated} considered the beam-squint and beam-split impacts in THz massive MIMO systems and designed an integrated sensing and communication scheme to improve the sensing accuracy and transmission rates. The authors in \cite{Wang2024Joint} optimized the design of hybrid precoder and radar receive beamforming for millimeter wave (mmWave)/Terahertz (THz) multi-user MIMO integrated sensing and communication (ISAC) systems. However, the methods in \cite{Zhang2022Time, Cong2023Vehicular,Ma2023Performance,Fan2023Radar,Yao2022Intelligent,Cheng2022Integrated,Mao2022Waveform,Chang2022Integrated,Gao2023Integrated,Chaccour2022Joint,Wu2023Sensing,Favarelli2023Sensor,Wang2024Joint} may not be able to capture the dynamics of vehicle network topologies caused by vehicle movements and dynamic wireless channels. In fact, vehicle network topology information can improve both sensing and communication services since it includes the connectivity information of all vehicles, which is crucial for managing the interference between THz sensing and communication links.  
\begin{table*}[ht]
  \renewcommand{\arraystretch}{1.2}
  \begin{center}
  \caption{Related Work on Communications and Sensing over Vehicular Networks.}
  \label{tab:table1}
  \resizebox{\textwidth}{!}{
  \begin{tabular}{|c|l|c|}   
  \hline
  \makebox[1cm][c]{\textbf{Reference}}& 
  \makebox[10cm][c]{\textbf{Main Contributions}} &  
  \makebox[3cm][c]{\textbf{Challenges}}
  \\ \hline
  \cite{Zhang2022Time,Yao2022Intelligent} & \makecell[l]{Time slot allocation for optimizing sensing and communications} & \multirow{6}{*}{\makecell[c]{Cannot be directly applied \\ for THz based sensing \\ and communications }}     \\   \cline{1-2} 
  \cite{Cong2023Vehicular} & \makecell[l]{Mobility aware sensing enabled beamforming optimization} &       
  \\  \cline{1-2}
  \cite{Ma2023Performance} & \makecell[l]{Cooperative detection in joint sensing and communication network} & 
  \\  \cline{1-2}
  \cite{Fan2023Radar} & \makecell[l]{Radar integrated MIMO communications for multi-hop networking}   &   
  \\   \cline{1-2}
  \cite{Cheng2022Integrated,Favarelli2023Sensor} & \makecell[l]{Tradeoff between sensing accuracy and communication throughput}  & 
  \\   \hline
  \cite{Mao2022Waveform,Wu2023Sensing} & \makecell[l]{Waveform design for mmWave/THz based sensing and communications}  & \multirow{5}{*}{\makecell[c]{Limited adaptability to \\ highly dynamic vehicle \\ network topologies}}  
  \\   \cline{1-2}
  \cite{Chaccour2022Joint} & \makecell[l]{THz enabled environmental sensing and data transmission for VR services}  & 
  \\   \cline{1-2}
  \cite{Chang2022Integrated} & \makecell[l]{Integrated sensing, communication, and control scheduling over THz bands}  & 
  \\  \cline{1-2}
  \cite{Gao2023Integrated,Wang2024Joint} & \makecell[l]{MmWave/THz beam management for integrated sensing and communications} & 
  \\   \hline
  \end{tabular}}
  \end{center}
\end{table*}

\begin{table*}[ht]
  \renewcommand{\arraystretch}{1.2}
  \begin{center}
  \caption{Related Work on GNNs.}
  \label{tab:table2}
  \resizebox{\textwidth}{!}{
  \begin{tabular}{|c|l|l|l|}   
  \hline
  \makebox[1.5cm][c]{\textbf{Reference}} & 
  \makebox[2.5cm][c]{\textbf{Graph Model}}& 
  \makebox[7cm][c]{\textbf{Graph Information}} & 
  \makebox[5.5cm][c]{\textbf{Application}}
  \\ \hline
  \cite{Gammelli2022Graph} & Recurrent GNN & Vehicle location, trajectory, and passenger information  & Vehicle scheduling 
  \\  \hline
  \cite{Jeon2020SCALE}& GCN & Vehicle location, velocity, angle, and connection & Vehicle trajectory prediction           \\ \hline
  \cite{Liu2023GADRL} & GAT & Vehicle location, direction, and computation tasks  & Computation task offloading   \\ \hline
  \cite{Zeng2023GNN} & Distributed GNN & Connection of mobile users & Cross-edge resource management                       \\ \hline
  \cite{Wang2023A} & GNN  & Channel conditions  and device connection  & RIS phase-shifts optimization \\ \hline
  \cite{Li2023Joint,Li2023Graph} & Heterogeneous GNN & Vehicle location, connection, and  interference & Sensing and communications optimization   
  \\ \hline
  \cite{Lee2022Intelligent}  & GCN & Vehicle connection & Sensing and communications optimization
  \\ \hline
  \cite{Zhang2023Cooperative} & Heterogeneous GNN & Communication interference and channel conditions & Trajectory optimization                  \\ \hline
  \cite{Kim2023A} & Bipartite GNN & Channel conditions and device connection  & Beamforming optimization   
  \\  \hline   
  \end{tabular}}
  \end{center}
\end{table*}

  A number of existing works such as in \cite{Lee2022Graph, Shen2023Graph, Gammelli2022Graph,Jeon2020SCALE, Liu2023GADRL, Zeng2023GNN, Wang2023A, Li2023Joint, Li2023Graph,Lee2022Intelligent, Zhang2023Cooperative, Kim2023A} studied the problem of using graph neural networks (GNNs) to extract topological and geographical location information of dynamic vehicle networks, as summarized in Table \uppercase\expandafter{\romannumeral2}. The authors in \cite{Lee2022Graph} and \cite{Shen2023Graph} provided a comprehensive survey of applying GNNs to extract representation vectors for mobile networks, and introduced the corresponding challenges, problems, and solutions. The authors in \cite{Gammelli2022Graph} solved a vehicle scheduling problem based on the topology features learned and extracted by a GNN. The authors in \cite{Jeon2020SCALE} used a graph convolutional network (GCN) to extract the vehicle topological and geographical information, including vehicle position, velocity, angle, and connection. In \cite{Liu2023GADRL}, a graph attention network (GAT) that can learn the importance coefficients of each neighboring vehicle was used to extract topological information of moving vehicles. In \cite{Zeng2023GNN}, the authors used a distributed GNN to handle an uncertain number of users and the constantly changing connections between them. The work in \cite{Wang2023A} used GNNs to capture the topological connection relations among multiple devices such as edge servers, reconfigurable intelligent surfaces (RISs), and edge users. The work in \cite{Li2023Joint} and \cite{Li2023Graph} developed a heterogeneous GNN based solution to extract the topology related features for different type of vehicles. The authors in \cite{Lee2022Intelligent} used GCNs to manage the joint radar-communication resources allocation in dynamic vehicle network topologies. In \cite{Zhang2023Cooperative}, the authors used GNNs to capture the communication interference and channel conditions among  UAVs and ground users. 
 The authors in \cite{Kim2023A} developed a scalable bipartite graph neural network (BGNN) that can quickly adapt to different system size (e.g., the number of antennas or users) so as to improve the efficiency of multi-antenna beamforming in a high dynamic environment.
 However, most of these works \cite{Gammelli2022Graph,Jeon2020SCALE, Liu2023GADRL, Zeng2023GNN, Wang2023A, Li2023Joint, Li2023Graph, Lee2022Intelligent, Zhang2023Cooperative, Kim2023A} used a single predefined GNN model to extract vehicle information from various network topologies and hence they did not consider whether the defined GNN model can process various vehicle network topologies thus reducing the information extracted by GNNs.

 The main contribution of this work is to design a novel framework that enables service provider vehicles (SPVs) to efficiently provide sensing and communication services to service request vehicles using THz bands. Our key contributions include:

 \begin{itemize}
    \item We consider a practical THz enabled joint sensing  and communication solution in vehicular networks, with which the SPVs provide sensing and communication services to the service request vehicles. To maximize the portion of successfully served vehicles, the SPVs need to choose their service mode to be providing sensing, communication, or both services.
     
    \item An optimization problem that aims to maximize the total number of successfully served vehicles by jointly determining the service mode of each SPV and the service request vehicles served by is introduced. The severe THz path loss, the blockages of sensing/communication links, as well as the environmental dynamics in regards to vehicle mobility and service needs, are all considered in this optimization problem.

    \item To solve this problem, we propose a dynamic GNN that selects appropriate graph information aggregation functions according to the vehicle network topology, thus extracting more vehicle network information compared to traditional static GNNs that use fixed aggregation functions for different vehicle network topologies. The aggregation function selection parameters and GNN model parameters are jointly optimized in the training process. Then, using the extracted vehicle network information, the service mode selection and service request vehicle connection of each SPV are determined.
 \end{itemize}
 Simulation results show that the proposed dynamic GNN based method can improve the number of successfully served vehicles by up to 17\% and 28\% compared to a GNN based algorithm with a fixed neural network model and a conventional optimization algorithm without using GNNs, respectively.


\section{System Model and Problem Formulation}
\label{sec:2}
 We consider a vehicular network that consists of a set $\mathcal{M}$ of $M$ communication service request vehicles, a set $\mathcal{N}$ of $N$ sensing service request vehicles, and a set $\mathcal{U}$ of $U$ SPVs, as shown in Fig. 1. In this network, each SPV uses the same range of THz bands to provide: 1) sensing services, 2) communication services, or 3) both sensing and communication services. We consider the vehicle network dynamics including vehicle movement, vehicle blockage, and unknown working state (i.e., providing services and cannot provide services) of each SPV. The movement of vehicles will change the vehicle locations, thereby altering the vehicle connection and wireless channels between vehicles. The vehicle blockage and unknown working state of each SPV will change the accessibility of each SPV, thereby increasing the dynamics of vehicle networks.
 Next, to model the accessibility of the SPVs, we introduce the vehicle blockage model, communication model, and sensing model. The main notations are summarized in Table \uppercase\expandafter{\romannumeral3}.

\begin{table}[t]
\renewcommand{\arraystretch}{1.2}
\centering
\caption{List of Main Notation}\label{tab:table3}
\begin{tabular}{|c||l|}   
   \hline
   \makebox[0cm][c]{\textbf{Notation}} & 
   \makebox[5cm][c]{\textbf{Description}}\\
   \hline
   $\mathcal{U}$  & The set of SPVs  \\   \hline
   $\mathcal{M}$ & The set of communication service request vehicles \\   \hline
   $\mathcal{N}$ & The set of sensing service request vehicles \\   \hline
   $\mathcal{B}$ & The set of successfully served communication vehicles \\ \hline
   $\mathcal{O}$ & The set of successfully served sensing vehicles \\   \hline
   $\mathcal{G}$ & A graph representation \\   \hline
   $\rho^{\textrm{C}}$ & Blockage indicator variable for communication mode \\   \hline
   $\rho^{\textrm{S}}$ & Blockage indicator variable for sensing mode \\   \hline
   $\omega_{u}$ & The state of SPV $u$ \\   \hline
   $\bm{\alpha}$ & Communication vehicle connection indicator matrix \\   \hline
   $\bm{\beta}$ & Sensing vehicle connection indicator matrix \\   \hline
   $\boldsymbol{w},\boldsymbol{\theta}$ &  Trainable parameters of GNN model \\   \hline 
   $H^{\textrm{B}}_{um}$ & Absorption loss between SPV $u$ and vehicle $m$ \\   \hline
   $H^{\textrm{F}}_{um} $ & Spreading loss between SPV $u$ and vehicle $m$  \\   \hline
   $A_{um}^{\textrm{T}}$ & The THz antenna gain of SPV $u$ serving vehicle $m$  \\  \hline
   $A_{mu}^{\textrm{R}}$ & The THz antenna gain of vehicle $m$ served by SPV $u$  \\   \hline  
   $Z_{um}^{\textrm{C}}(\bm{\alpha},\bm{\beta})$ & The interference of vehicle $m$ served by SPV $u$ \\   \hline
   $Z_{um}^{\textrm{S}}(\bm{\alpha},\bm{\beta})$ & The interference of vehicle $n$ served by SPV $u$  \\   \hline   
   $\mathrm{\lambda}_{um}^{\textrm{C}}(\bm{\alpha},\bm{\beta})$ & The SINR of vehicle $m$ served by SPV $u$  \\   \hline
   $\mathrm{\lambda}_{um}^{\textrm{S}}(\bm{\alpha},\bm{\beta})$ & The SINR of vehicle $n$ served by SPV $u$\\   \hline  $E_{um}^{\textrm{C}}\left(\bm{\alpha},\bm{\beta}\right)$ &  Data rate of SPV $u$ transmitting data to vehicle $m$ \\   \hline
   $D_{max}$ & Maximum tolerable delay of communication services \\   \hline
   $\mathrm{\lambda}_{min}$ &  Minimum SINR requirement of sensing services \\   \hline
   
\end{tabular}
\end{table}

\subsection{Vehicle Blockage Model}
The projection of a building or tree on the ground is modeled as a quadrilateral. The transmission link between SPV $u$ and communication service request vehicle $m$ will be blocked when the line segment between SPV $u$ and communication service request vehicle $m$ intersects with one of diagonals of the quadrilateral. Specifically, if 1) SPV $u$ and communication service request vehicle $m$ are on the different sides of the diagonal, 2) the vertices of the diagonal are on the different sides of the straight line passing through SPV $u$ and communication service request vehicle $m$, the connection between SPV $u$ and vehicle $m$ will be blocked. Here, a binary variable $\rho_{um}^{\textrm{C}}$ that indicates whether a blockage exists between SPV $u$ and communication service request vehicle $m$ is expressed as

\begin{equation}
\label{blockage}
\rho_{um}^{\textrm{C}} = \left\{
\begin{array}{cl}
0, &  \text{if a blockage exists between vehicle $u$ and $m$ } ,\\
1, & \text{otherwise},
 \end{array}
\right.
\end{equation}
within which $\rho_{um}^{\textrm{C}} = 1$ means that the communication link between SPV $u$ and communication service request vehicle $m$ is  line-of-sight (LoS); otherwise, we have $\rho_{um}^{\textrm{C}} = 0$. Similarly, the link between SPV $u$ and sensing service request vehicle $n$ will be blocked when the line segment between SPV $u$ and sensing service request vehicle $n$ intersects with one of diagonals of the quadrilateral. The binary variable $\rho_{un}^{\textrm{S}}$ that indicates whether a blockage exists between SPV $u$ and sensing service request vehicle $n$ is expressed as
\begin{equation}
\label{blockage2}
\rho_{un}^{\textrm{S}} = \left\{
\begin{array}{cl}
0, &  \text{if a blockage exists between vehicle $u$ and $n$} ,\\
1, & \text{otherwise},
 \end{array}
\right.
\end{equation}
 within which $\rho_{un}^{\textrm{S}} = 1$ means that the sensing link between SPV $u$ and sensing service request vehicle $n$ is LoS; otherwise, we have $\rho_{un}^{\textrm{S}} = 0$.  In (\ref{blockage}) and (\ref{blockage2}), we have considered the  buildings and trees that may cause the blockage between SPVs and service request vehicles. In future, we will study how terrain affects the performance of joint sensing and communications in vehicular networks based on satellite map.

\begin{figure}[t]
\centering
\setlength{\abovecaptionskip}{-0.45cm} 
\setlength{\belowcaptionskip}{-0.45cm}
\includegraphics[width=0.9\linewidth]{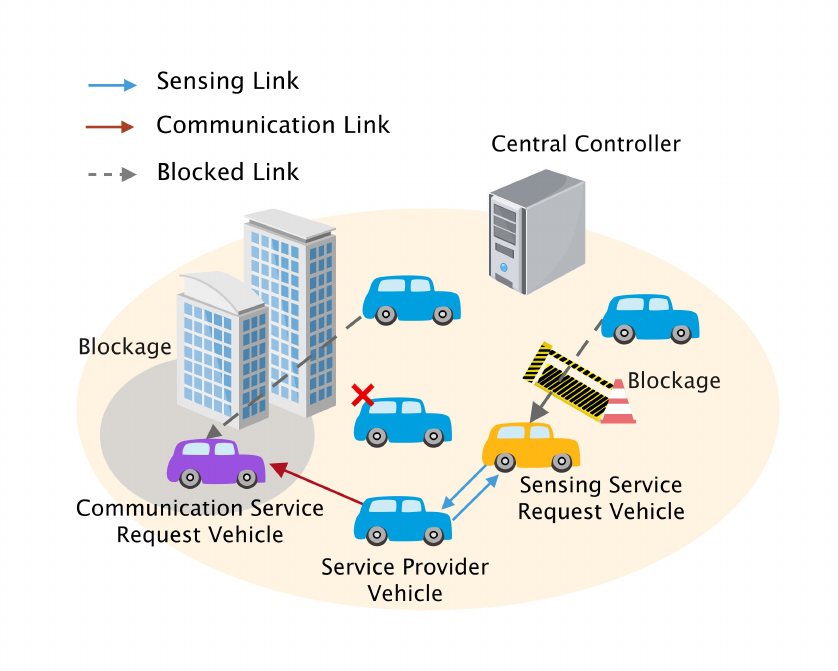}
\caption{Illustration of the considered vehicular network model.}
\label{system_model}
\end{figure}

\subsection{Communication Model}
We consider a practical vehicle network where each SPV may not always function, such that each SPV has two states: 1) active state at which the SPV can provide services and 2) deactivate state at which the SPV cannot provide any service. Let $\omega_{u}$ be the state of an SPV with $\omega_{u}=1$ indicating that vehicle $u$ can provide sensing, communication, or both services; otherwise, we have $\omega_{u}=0$. The probability of an SPV in the active state is $p$.

The power transmitted by SPV $u$ and received by communication service request vehicle $m$ is 
\begin{equation}
\begin{aligned}\label{power}
S_{um}=\frac{\omega_{u} \rho_{um}^{\textrm{C}} P_{u}A_{um}^{\textrm{T}} A_{mu}^{\textrm{R}}}{H^{\textrm{B}}_{um}H^{\textrm{F}}_{um}},
\end{aligned}
\end{equation}
within which $P_{u}$ is the transmit power of SPV $u$, $H^{\textrm{F}}_{um}=\frac{\left(4\pi f d_{um}\right)^2}{c^2}$ is the free space path gain, $H^{\textrm{B}}_{um}=\frac{1}{r\left(d_{um}\right)}$ is the molecular absorption path gain with $r\left(d_{um}\right) \approx e^{-\tau\left(f\right) d_{um}}$ being the transmittance of the medium, $\tau\left(f\right)$ is the overall absorption coefficient of the medium, $d_{um}$ is the distance between SPV $u$ and communication service request vehicle $m$, $c$ is the speed of light, and $f$ is the operating frequency. Since the molecular absorption loss $H^{\textrm{B}}_{um}$ and free space path loss $H^{\textrm{F}}_{um}$ can cause the severe attenuation, higher antenna gains are required in the THz bands so as to compensate the severe path loss. The effective THz antenna gain can be represented as a function of the horizontal and vertical beamwidths. Let $A_{um}^{\textrm{T}}$ represent the effective antenna gain of SPV $u$ transmitting data to communication service request vehicle $m$,
which can be denoted by $A_{um}^{\textrm{T}}=\frac{4\pi}{(\iota+1) \Gamma_{\varrho_{u},\varsigma_{u}}}$ for the main lobe
and $A_{um}^{\textrm{T}}=\frac{4\pi \iota}{\left(\iota+1\right)\left(4\pi-\Gamma_{\varrho_{u},\varsigma_{u}}\right)}$ for the side lobes. Similarly, $A_{mu}^{\textrm{R}}$ represents the effective antenna gain of communication service request vehicle $m$ served by SPV $u$, which can be denoted by $A_{mu}^{\textrm{R}}=\frac{4\pi}{(\iota+1) \Gamma_{\varrho_{m},\varsigma_{m}}}$ for the main lobe
and $A_{mu}^{\textrm{R}}=\frac{4\pi \iota}{\left(\iota+1\right)\left(4\pi-\Gamma_{\varrho_{m},\varsigma_{m}}\right)}$ for the side lobes. $\Gamma_{\varrho_{u},\varsigma_{u}}=4 \arcsin \left(\tan \left(\frac{\varrho_{u}}{2}\right) \tan \left(\frac{\varsigma_{u}}{2}\right)\right)$, within which $\varrho_{u}$ and $\varsigma_{u}$ represent, respectively, the horizontal and vertical beamwidths of the antenna of SPV $u$. $\iota$ captures the side lobe power to main lobe power ration. Here, if the horizontal and vertical emission angles of SPV $u$ towards its associated vehicle $m$ are within the horizontal and vertical beamwidths of SPV $u$, the effective THz antenna gain of main lobe is selected for SPV $u$; otherwise, the effective THz antenna gain of side lobes is selected for SPV $u$.

The interference at communication service request vehicle $m$ served by SPV $u$ is
\begin{equation}\label{eq:int_communication}
\begin{split}
Z_{um}^{\textrm{C}}\left(\bm{\alpha},\bm{\beta}\right)=\sum_{i\in  \mathcal{U}\setminus\{u\}}\sum_{m^{\prime}\in  \mathcal{M}}\frac {\omega_{i} \rho_{im^{\prime}}^{\textrm{C}} \rho_{im}^{\textrm{C}} \alpha_{im^{\prime}} P_{i} A_{im}^{\textrm{T}} A_{mi}^{\textrm{R}} }{H^{\textrm{B}}_{im}H^{\textrm{F}}_{im}}\\
+\sum_{i\in  \mathcal{U}\setminus\{u\}}\sum_{n^{\prime}\in  \mathcal{N}}\frac {  \omega_{i} \rho_{in^{\prime}}^{\textrm{S}} \rho_{im}^{\textrm{C}} \beta_{in^{\prime}} P_{i} A_{im}^{\textrm{T}} A_{mi}^{\textrm{R}} }{H^{\textrm{B}}_{im}H^{\textrm{F}}_{im}}\\
+ \sum_{n^{\prime}\in  \mathcal{N}}\frac {  \omega_{u} \rho_{un^{\prime}}^{\textrm{S}} \rho_{um}^{\textrm{C}} \beta_{un^{\prime}} P_{u} A_{um}^{\textrm{T}} A_{mu}^{\textrm{R}} }{H^{\textrm{B}}_{um}H^{\textrm{F}}_{um}},
\end{split}
\end{equation}
within which $\bm{\alpha}=\left[\bm{\alpha}_{1}, \cdots, \bm{\alpha}_{M}\right]$ is communication service vehicle connection indicator matrix with $\bm{\alpha}_{m}=\left[{\alpha}_{1m}, \cdots, {\alpha}_{Um}\right]$, and $\bm{\beta}=\left[\bm{\beta}_{1}, \cdots, \bm{\beta}_{N}\right]$ is sensing service vehicle connection indicator matrix with $\bm{\beta}_{n}=\left[{\beta}_{1n}, \cdots, {\beta}_{Un}\right]$. $\alpha_{im}=1$ represents that SPV $i$ is selected to serve communication service request vehicle $m$ in the communication mode; otherwise, $\alpha_{im}=0$. Similarly, $\beta_{in}=1$ represents that SPV $i$ is selected to detect sensing service request vehicle $n$ in the sensing mode; otherwise, $\beta_{in}=0$. 
Note that the first two terms of (\ref{eq:int_communication}) capture, respectively, the interference caused by other communication services, and by other sensing services. The third term is the interference caused by the sensing transmitting antenna of the current SPV since each SPV can simultaneously provide sensing and communication services. Here, we ignored the sensing interference caused by communication service provided by the same vehicle since we can optimize the antenna array to eliminate the interference, as done in \cite{Liu2022Optimal} and \cite{Nuss2017MIMO}. 

The signal-to-interference-plus-noise ratio (SINR) of communication service request vehicle $m$ served by SPV $u$ is 
\begin{equation}
\begin{aligned}
\mathrm{\lambda}_{um}^{\textrm{C}}\left(\bm{\alpha},\bm{\beta}\right)=\frac{ S_{um}} {Z_{um}^{\textrm{C}}\left(\bm{\alpha},\bm{\beta}\right)+\varepsilon_{um}},
\end{aligned}
\end{equation}
where $\varepsilon_{um}= \sum_{i\in  \mathcal{U}\setminus\{u\}}  \omega_i \rho_{im}^{\textrm{C}} P_{i} A_{im}^{\textrm{T}} A_{mi}^{\textrm{R}} \left(1-r\left(d_{im}\right)\right)/H^{\textrm{F}}_{im}\\+\varepsilon_0$ with $\varepsilon_0$ being the Johnson-Nyquist noise power. $\varepsilon_{um}$ is caused by thermal agitation of electrons and molecular absorption.

Therefore, the data rate of the link between SPV $u$ and the communication service request vehicle $m$ is
\begin{equation}
\begin{aligned}
E_{um}^{\textrm{C}}\left(\bm{\alpha},\bm{\beta}\right) = B \log_2 \left(1+\mathrm{\lambda}_{um}^{\textrm{C}}\left(\bm{\alpha},\bm{\beta}\right)\right),
\end{aligned}
\end{equation}
with $B$ being the bandwidth.

\subsection{Sensing Model}
The interference at sensing service request vehicle $n$ served by SPV $u$ is
\begin{equation}\label{eq:int_sensing}
\begin{split}
Z_{un}^{\textrm{S}}\left(\bm{\alpha},\bm{\beta}\right)=\sum_{i\in  \mathcal{U}\setminus\{u\}} \sum_{m^{\prime} \in  \mathcal{M}} \frac { \omega_i \rho_{im^{\prime}}^{\textrm{C}} \rho_{iu}^{\textrm{S}} \alpha_{im^{\prime}} P_{i} A_{iu}^{\textrm{T}} A_{ui}^{\textrm{R}} }{H^{\textrm{B}}_{iu}H^{\textrm{F}}_{iu}}  \\
+\sum_{i\in  \mathcal{U}\setminus\{u\}} \sum_{n^{\prime} \in  \mathcal{N}} \frac {  \omega_i \rho_{in^{\prime}}^{\textrm{S}} \rho_{iu}^{\textrm{S}} \beta_{in^{\prime}} P_{i} A_{iu}^{\textrm{T}} A_{ui}^{\textrm{R}} }{H^{\textrm{B}}_{iu}H^{\textrm{F}}_{iu}}\\
+ \sum_{i\in  \mathcal{U}\setminus\{u\}} \sum_{m^{\prime} \in  \mathcal{M}}  \frac { \omega_i \rho_{im^{\prime}}^{\textrm{C}} \rho_{in}^{\textrm{S}}\alpha_{im^{\prime}} P_{i} A_{in}^{\textrm{T}} A_{nu}^{\textrm{R}} \kappa_{in} c^2}{(4 \pi)^{3} f^2 d_{in}^{2} d_{un}^{2} H^{\textrm{B}}_{in}H^{\textrm{B}}_{un}}\\
+ \sum_{i\in  \mathcal{U}\setminus\{u\}} \sum_{n^{\prime} \in  \mathcal{N}}  \frac { \omega_i \rho_{in^{\prime}}^{\textrm{S}} \rho_{in}^{\textrm{S}} \beta_{in^{\prime}} P_{i} A_{in}^{\textrm{T}} A_{nu}^{\textrm{R}} \kappa_{in} c^2}{(4 \pi)^{3} f^2 d_{in}^{2} d_{un}^{2} H^{\textrm{B}}_{in}H^{\textrm{B}}_{un}},
\end{split}
\end{equation}
within which $\kappa_{in}$ is the radar cross section when SPV $i$ provides sensing service for vehicle $n$. In (\ref{eq:int_sensing}), the first term indicates the interference caused by other SPVs providing communication services with line-of-sight transmission links. The second term indicates the interference caused by other SPVs providing sensing services with line-of-sight transmission links. The third term indicates the interference caused by other SPVs providing communication services via scattering paths. The last term indicates the interference caused by other SPVs providing sensing services via scattering paths. From (\ref{eq:int_communication}) and (\ref{eq:int_sensing}), we see that a sensing service request vehicle is interfered by the scattering path interference caused by other service request vehicles. However, the scattering path interference will not interfere communication service request vehicles since sensing services are more sensitive to scattered sensing signals \cite{Zhang2022Time}. 

Given (\ref{eq:int_sensing}), the SINR of sensing service request vehicle $n$ served by SPV $u$ is
\begin{equation}
\begin{aligned}
\mathrm{\lambda}_{un}^{\textrm{S}}\left(\bm{\alpha},\bm{\beta}\right)=\frac { P_{u} A_{un}^{\textrm{T}} A_{nu}^{\textrm{R}}  (H^{\textrm{S}}_{un})^{-1} (H^{\textrm{B}}_{un})^{-1} }{Z_{un}^{\textrm{S}}\left(\bm{\alpha},\bm{\beta}\right)+\varepsilon_{un}},
\end{aligned}
\end{equation}
within which $H^{\textrm{S}}_{un}=\frac {(4 \pi)^{3} f^{2} d_{un}^{4}} {\kappa_{un} c^2}$ is the spreading loss of the sensing signal sent by SPV $u$, reflected by sensing service request vehicle $n$, and then received by SPV $u$.

\subsection{Successfully Served Vehicles}
A successfully served vehicle must have its communication or sensing service requirement satisfied. Let $Q_m$ be the size of the information requested by communication service request vehicle $m$, the transmission delay between communication service request vehicle $m$ and its associated SPV is $\frac{Q_m}{\sum_{u\in  \mathcal{U}}  \alpha_{um} E_{um}^{\textrm{C}}\left(\bm{\alpha},\bm{\beta}\right)}$. Then, the set of successfully served communication service request vehicles is given by 
\begin{equation}
\begin{aligned}\label{eq:req_com}
{\mathcal{B}}=\{m|\frac{Q_m}{\sum_{u\in  \mathcal{U}}  \alpha_{um} E_{um}^{\textrm{C}}\left(\bm{\alpha},\bm{\beta}\right)} \leq D_{max}, \forall m \in {\mathcal{M}}, \forall u \in {\mathcal{U}} \},
\end{aligned}
\end{equation}
within which $D_{\max}$ is the maximum tolerable delay of the communication service. The set of successfully served sensing service request vehicles is 
\begin{equation}
\begin{aligned}\label{eq:req_sen}
{\mathcal{O}}=\{n| \sum_{u\in  \mathcal{U}} \beta_{un} \mathrm{\lambda}_{un}^{\textrm{S}}\left(\bm{\alpha},\bm{\beta}\right) \geq \mathrm{\lambda}_{min}, \forall n \in {\mathcal{N}}, \forall u \in {\mathcal{U}}\},
\end{aligned}
\end{equation}
within which $\mathrm{\lambda}_{min}$ is the minimum SINR threshold required by the sensing service.

\subsection{Problem Formulation}
In the defined system model, the goal is to optimize the service mode (i.e., providing sensing, communication, or both services) of each SPV and its associated vehicles to maximize the number of successfully served vehicles, which is formulated as optimization problem
\begin{subequations}\label{eq:litdiff}
\begin{align}
	\mathop{\mbox{max}}_{\bm{\alpha},\bm{\beta}} \quad
&   { {|\mathcal{B}|}+{|\mathcal{O}|}}\tag{\ref{eq:litdiff}}\\
	\mbox{s.t.} 
    \quad
&(\ref{eq:req_com}) - (\ref{eq:req_sen}), \label{eq:11a} \\
    &\sum_{u\in  \mathcal{U}}\alpha_{um}=1,\alpha_{um} \in \{0,1\},\forall m \in \mathcal{M}, \forall u \in {\mathcal{U}},\label{eq:11b} \\
	&\sum_{u\in  \mathcal{U}}\beta_{un}=1,\beta _{un} \in \{0,1\}, \forall n \in \mathcal{N}, \forall u \in {\mathcal{U}},\label{eq:11c} 
\end{align}
\end{subequations}
 within which ${|\mathcal{B}|}$ is the number of successfully served communication service request vehicles, and ${|\mathcal{O}|}$ is the number of successfully served sensing service request vehicles.  In (\ref{eq:litdiff}), constraint (\ref{eq:11b}) requires a communication service request vehicle to be served by only one SPV. Constraint (\ref{eq:11c}) requires a sensing service request vehicle to be served by only one SPV. 

Problem (\ref{eq:litdiff}) is challenging to solve for multiple reasons. First, problem (\ref{eq:litdiff}) is non-convex, which means applying the traditional optimization algorithms to solve problem (\ref{eq:litdiff}) can incur significantly high complexity. Second, the objective function, as well as the constraints in (11a) are all unknown and change in some unknown pattern with the dynamic vehicle network topology. The traditional optimization methods may not be applied for dynamic vehicle network topologies caused by vehicle movement, vehicle blockage, and unknown working state (i.e., providing services and cannot provide services) of each SPV. When the vehicle network topology changes, the traditional optimization methods need to be executed again to re-optimize service mode selection and service request vehicle connection. Third, the vehicle dynamics caused by THz bands make the optimization problem (11) hard to solve. Compared to other frequency bands such as mmWave, using THz bands for providing both communication and sensing services in vehicle networks requires the designed algorithm to be quickly adapted to the vehicle network dynamics caused by blockages of THz communications and sensing. In consequence, the methods designed for mmWave based communications and sensing may not be used for THz bands, whereas the proposed method can also be used for mmWave bands. To solve this problem, we use GNNs to extract not only geographical location information (e.g., vehicle location) but also topological information (e.g., vehicle connection and vehicle interference). 
Compared to traditional optimization methods and other neural network based algorithms that need to be retrained when vehicle topologies change (e.g, adding new vehicles or removing existing vehicles), a trained dynamic GNN model can be used to extract the features of newly arrived vehicles without retraining thus reducing training complexity.

\section{Dynamic GNN based Solution}
\label{sec:3}

In this section, a dynamic GNN based algorithm is introduced to solve problem (\ref{eq:litdiff}).  Different from the static GNN model \cite{Li2023Joint} that uses fixed node aggregation functions to extract vehicle network information for different vehicle network topologies, the proposed dynamic GNN model can select appropriate graph information aggregation functions from eight types of node aggregation functions and three types of layer aggregation functions. Hence, the proposed dynamic GNN based algorithm can extract more vehicle topology features and find a better service mode for each SPV and its associated service request vehicles. Next, we first introduce graph representation of the considered vehicular networks. Then, we present the components of the GNN based algorithm and the training process. Finally, we provide the entire procedure of using the proposed algorithm to select vehicle service mode and determine the service request vehicle connection.

\subsection{Representation of Vehicular Networks via Graphs }
\begin{figure}[t]
\centering
\setlength{\abovecaptionskip}{-0.3cm} 
\setlength{\belowcaptionskip}{-0.45cm}
\includegraphics[width=0.9\linewidth]{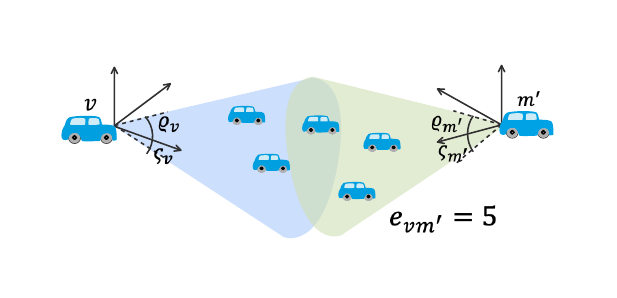}
\caption{Visualization of node feature.}
\label{8-feature}
\end{figure}

Here, we explain the use of graphs to represent the considered vehicular network. Each vehicle is modeled as a node while each connected link (e.g., sensing link, communication link, and interference link) between two vehicles is modeled as an edge. Let $\mathcal{G} = \left(\mathcal{V}, \mathcal{E}\right)$ represent a graph with node features $\bm{f} \in \mathbb{R}^{ P \times {|\mathcal{V}|}}$, within which $\mathcal{V}$ and $\mathcal{E}$ represent the node and edge sets, respectively. The node set ${\mathcal{V}} = \mathcal{U} \cup \mathcal{M} \cup \mathcal{N}$ contains three types of vehicles, ${|\mathcal{V}|=U+M+N}$ is the number of vehicles, and $P = U+2(M+N)$ is the dimension of node features. Specifically, the node features can be defined by $\bm{f}=\left[\bm{f}_{1}, \cdots, \bm{f}_{V}\right]$ with $\boldsymbol{f}_{v}=\left[e_{v1},\dots,e_{vM^{\prime}}, g_{v1}, \dots, g_{vV}\right]^\top$ being the node feature for vehicle $v \in {\mathcal{V}}$, within which $e_{vm^{\prime}}$ captures the number of SPVs that fall within the LoS link between vehicle $v$ and $m^{\prime}$ (as shown in Fig. \ref{8-feature}), ${g}_{vv^{\prime}}=\left(H^{\textrm{B}}_{vv^{\prime}}H^{\textrm{F}}_{vv^{\prime}}\right)^{-1}$ captures the free space and molecular absorption path gain between vehicle $v$ and $v^{\prime}$, and ${\mathcal{M^{\prime}}}=\mathcal{M} \cup \mathcal{N}$ is the set of service request vehicles. Moreover, we assume that there is an edge between SPV $u$ and service request vehicle $m^{\prime}$ when $\omega_{u} \rho_{u{m^{\prime}}}=1$, which guarantees that SPV $u$ is working in an active state and the connected link between SPV $u$ and  service request vehicle
$m^{\prime}$ is not blocked. Since the locations and working state of vehicles vary over time, the edge set $\mathcal{E}$ will change dynamically in different vehicle network topologies.

\begin{figure}[t]
\centering
\setlength{\abovecaptionskip}{-0.3cm} 
\setlength{\belowcaptionskip}{-0.45cm}
\includegraphics[width=0.9\linewidth]{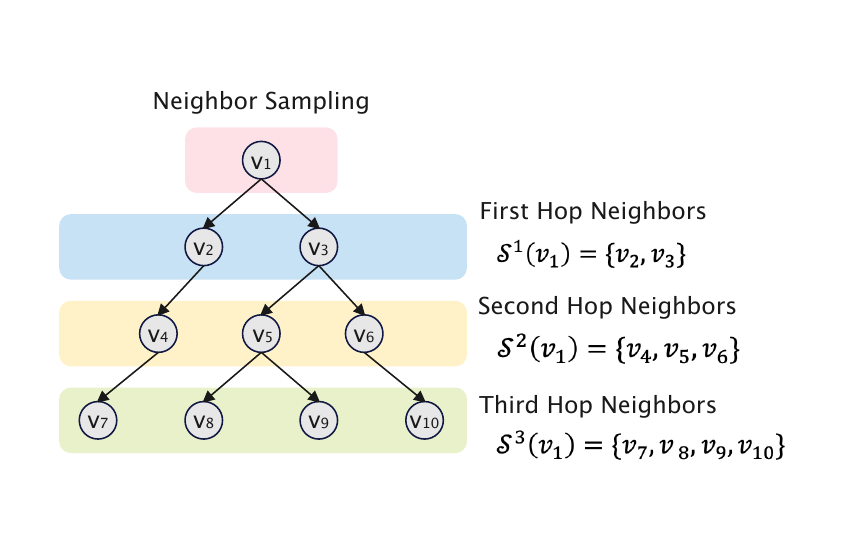}
\caption{Neighbor sampling process of vehicle $v$.}
\label{Sampling}
\end{figure}

 For each vehicle $v \in {\mathcal{V}}$, we define three types of vehicles as follows: 1) the first hop vehicles which can directly connect to vehicle $u$ are represented by ${\mathcal{S}}^1\left(v\right)= \{v^{\prime} \in {\mathcal{V}}|\left(v,v^{\prime}\right) \in {\mathcal{E}}\}$ with $|{\mathcal{S}}^1(v)|$ being the number of vehicles in ${\mathcal{S}}^1\left(v\right)$, 2)the second hop vehicles that connect to vehicle $v$ through an intermediate vehicle (i.e. a fist hop vehicle) are captured in set ${\mathcal{S}}^2\left(v\right)=\{{\mathcal{S}}^1\left(v^{\prime}\right)|v^{\prime} \in {\mathcal{S}}^1\left(v\right)\}$ with $|{\mathcal{S}}^2(v)|$ being the number of vehicles in ${\mathcal{S}}^2\left(v\right)$, and 3) the third hop vehicles that connect to vehicle $v$ through two intermediate vehicles (i.e. the first and second hop vehicles) are captured in set ${\mathcal{S}}^3\left(v\right)=\{{\mathcal{S}}^1\left(v^{\prime}\right)| v^{\prime} \in {\mathcal{S}}^2\left(v\right)\}$ with $|{\mathcal{S}}^3(v)|$ being the number of vehicles in ${\mathcal{S}}^3\left(v\right)$. For the example shown in Fig.~\ref{Sampling}, we have ${\mathcal{S}}^1\left(v_1\right)=\left\{v_2, v_3\right\}$, ${\mathcal{S}}^2\left(v_1\right)=\{v_4, v_5, v_6\}$, and ${\mathcal{S}}^3\left(v_1\right)=\{v_7, v_8, v_9, v_{10}\}$. Here,  we consider three hop vehicles because the information of three hop vehicles can represent the vehicle connection and topological information of a given vehicle.

\begin{figure*}[t]
\centering
\setlength{\abovecaptionskip}{-0.3cm} 
\setlength{\belowcaptionskip}{-0.45cm}
\vspace{-0cm}
\includegraphics[width=0.9\linewidth]{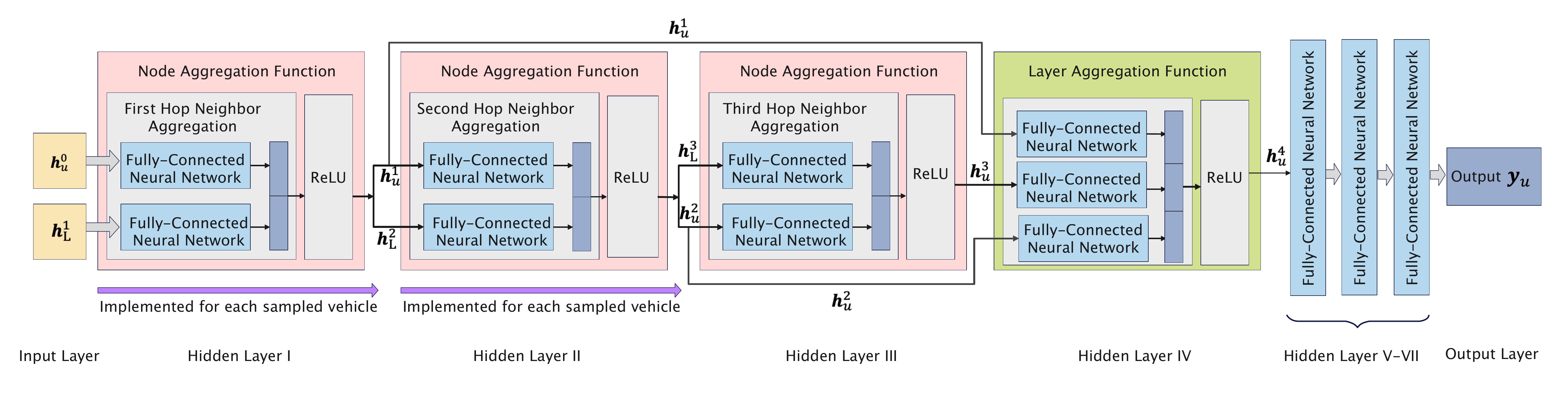}
\caption{Structure of the proposed dynamic GNN model.}
\label{nn}
\end{figure*}

\subsection{Components of the GNN based Algorithm}
\begin{figure*}[t]
\centering   
\includegraphics[width=0.9\linewidth]{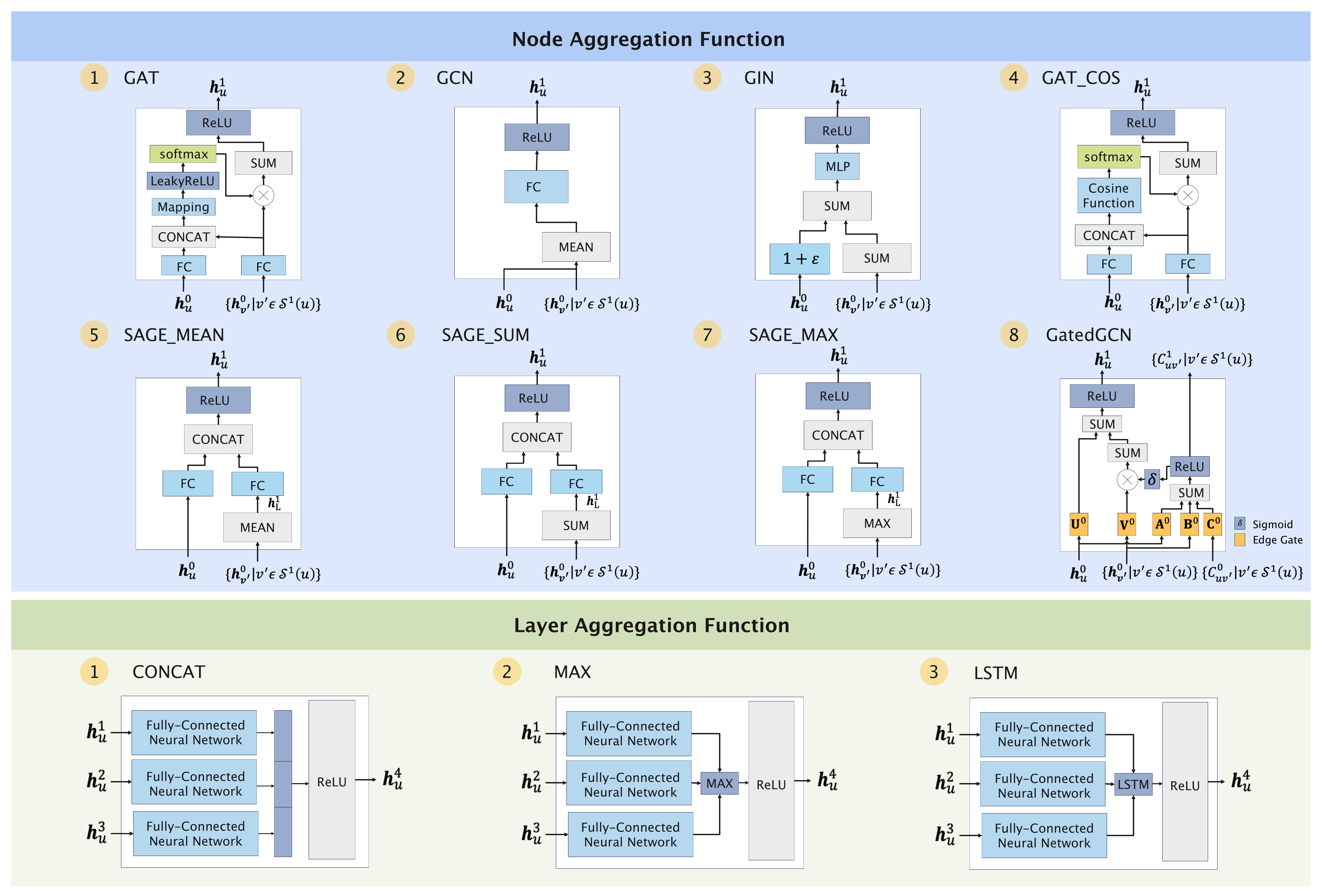}
\caption{Key explanations to the node and layer aggregation functions.}
\label{8-agg}
\end{figure*}

As shown in Fig.~\ref{nn}, the components of the proposed dynamic GNN based algorithm that is implemented by a central controller are \cite{Zhao2021Search}: 1) input layer, 2) hidden layer \uppercase\expandafter{\romannumeral1}, 3) hidden layer \uppercase\expandafter{\romannumeral2}, 4) hidden layer \uppercase\expandafter{\romannumeral3}, 5) hidden layer \uppercase\expandafter{\romannumeral4}, 6) hidden layer \uppercase\expandafter{\romannumeral5}-\uppercase\expandafter{\romannumeral7}, and 7) output layer. These components are specified as follows:
\begin{itemize}


    \item \emph{Input}:
   The GNN input layer is connected to two paralleled fully connected layers (FCLs). The input of the first fully connected layer is $\boldsymbol{h}_u^0=\boldsymbol{f}_u  \in \mathbb{R}^{P \times 1}$ and the input of the second fully connected layer is $\boldsymbol{h}_{\textrm{L}}^{1} \in \mathbb{R}^{P \times 1}$, where   
    \begin{equation}
    \label{input2-1}   \boldsymbol{h}_{\textrm{L}}^{1}=\frac{1}{|{\mathcal{S}}^1\left(u\right)|}\sum_{v^{\prime} \in {\mathcal{S}}^1\left(u\right)}{\boldsymbol{h}_{v^{\prime}}^{0}},
    \end{equation}
    with ${\boldsymbol{h}_{v^{\prime}}^{0}}=\boldsymbol{f}_{v^{\prime}} \in \mathbb{R}^{P \times 1}$.
    
    \item \emph{Hidden Layer \uppercase\expandafter{\romannumeral1}}: Hidden layer I contains two paralleled FCLs. It is in charge of the graph information extraction for each each vehicle $u$'s first hop vehicles, and outputs
     \begin{equation}
     \label{con1}
      {\boldsymbol{h}_{u}^{1}} = \sigma\left( \left[{\boldsymbol{w}_1}{\boldsymbol{h}_{u}^{0}} \|
      {\boldsymbol{w}_2} {\boldsymbol{h}^{1}_{\textrm{L}}}
      \right]\right), 
     \end{equation}
    within which $\sigma\left(\cdot\right)$ is the rectified linear unit function, $\cdot \| \cdot$ is the vector concatenation.  $\boldsymbol{w}_1 \in \mathbb{R}^{\left(\Omega_{0}/2 \right) \times P}$ and $\boldsymbol{w}_2 \in \mathbb{R}^{\left(\Omega_{0}/2 \right) \times P}$ are, respectively, the weight parameters of the two FCLs, with $\Omega_{0}$ being the size of the graph information vector. Here, (\ref{input2-1}) and (\ref{con1}) are node aggregation functions that extract the graph information of vehicle $u$. We can also consider other types of node aggregation functions \cite{Zhao2021Search} as shown in Fig. \ref{8-agg}. 
    Notice that, (\ref{input2-1}) and (\ref{con1}) can only extract the graph information for each vehicle once a time, while the graph information $\boldsymbol{h}_{v^{\prime}}^{1}$ of all vehicles is required for solving the joint model selection and vehicle connection problem, the implementation of (\ref{input2-1}) and (\ref{con1}) needs to repeated for $|{\mathcal{S}}^1(u)|$ times.
    
  
    \item \emph{Hidden Layer  \uppercase\expandafter{\romannumeral2}}: Hidden layer II also consists of two paralleled FCLs, with the input of the first fully connected layer being $\boldsymbol{h}_u^1  \in \mathbb{R}^{\Omega_{0} \times 1}$ and the input of the second fully connected layer being $\boldsymbol{h}_{\textrm{L}}^{2} \in \mathbb{R}^{\Omega_{0}\times 1}$, with   
    \begin{equation}
    \label{agg2-1}
    \boldsymbol{h}_{\textrm{L}}^{2}=\frac{1}{|{\mathcal{S}}^1\left(u\right)|}\sum_{v^{\prime} \in {\mathcal{S}}^1 \left(u\right)}{\boldsymbol{h}_{v^{\prime}}^{1}}.
    \end{equation}
   The hidden layer II extracts the graph information for each vehicle $u$'s second hop vehicles, and outputs
    \begin{equation}
    \label{agg2-2}
      {\boldsymbol{h}_{u}^{2}} = \sigma\left( \left[{\boldsymbol{w}_3}{\boldsymbol{h}_{u}^{1}} \|
      {\boldsymbol{w}_4} {\boldsymbol{h}^{2}_{{\textrm{L}}}}
      \right]\right), 
     \end{equation}
     within which $\boldsymbol{h}_u^2 \in   \mathbb{R}^{\Omega_{0} \times 1}$, $\boldsymbol{w}_3 \in \mathbb{R}^{\left(\Omega_{0}/2 \right) \times \Omega_{0}}$ and $\boldsymbol{w}_4 \in \mathbb{R}^{\left(\Omega_{0}/2 \right) \times \Omega_{0}}$ are the weight parameters of the two FCLs, respectively. (\ref{agg2-1}) and (\ref{agg2-2}) also constitute a node aggregation function that extracts the graph information of vehicle $u$ and its second hop neighbors, and need to be implemented for $|{\mathcal{S}}^1(u)|$ times to obtain the the graph information $\boldsymbol{h}_{v^{\prime}}^{2}$ of all sampled vehicles in ${\mathcal{S}}^1 \left(u\right)$. 
     

    \item \emph{Hidden Layer  \uppercase\expandafter{\romannumeral3}}: This layer contains two paralleled FCLs, with the input of the first fully connected layer being $\boldsymbol{h}_u^2  \in \mathbb{R}^{\Omega_{0} \times 1}$ and the input of the second fully connected layer being $\boldsymbol{h}_{\textrm{L}}^{3} \in \mathbb{R}^{\Omega_{0} \times 1}$, within which  
    \begin{equation}
    \label{agg3-1}
    \boldsymbol{h}_{\textrm{L}}^{3}=\frac{1}{|{\mathcal{S}}^1\left(u\right)|}\sum_{v^{\prime} \in {\mathcal{S}}^1 \left(u\right)}{\boldsymbol{h}_{v^{\prime}}^{2}}.
    \end{equation}
    Hidden Layer  \uppercase\expandafter{\romannumeral3} takes charge of the graph information extraction of each vehicle $u$'s third hop vehicles, and outputs
    \begin{equation}
    \label{agg3-2}
      {\boldsymbol{h}_{u}^{3}} = \sigma\left( \left[{\boldsymbol{w}_5}{\boldsymbol{h}_{u}^{2}} \|
      {\boldsymbol{w}_6} {\boldsymbol{h}^{3}_{{\textrm{L}}}}
      \right]\right), 
     \end{equation}
      within which $\boldsymbol{h}_u^3 \in   \mathbb{R}^{\Omega_{0} \times 1}$, $\boldsymbol{w}_5 \in \mathbb{R}^{\left(\Omega_{0}/2 \right) \times \Omega_{0}}$ and $\boldsymbol{w}_6 \in \mathbb{R}^{\left(\Omega_{0}/2 \right) \times \Omega_{0}}$ are the weight parameters of the two FCLs, respectively. Similarly, (\ref{agg3-1}) and (\ref{agg3-2}) also constitute a node aggregation function that extracts the graph information of vehicle $u$ and its third hop neighbors.
    
    \item \emph{Hidden Layer \uppercase\expandafter{\romannumeral4}}: This layer contains three paralleled FCLs and it  combines the output of hidden layers \uppercase\expandafter{\romannumeral1}-\uppercase\expandafter{\romannumeral3} (i.e., ${\boldsymbol{h}_{u}^{1}}$, ${\boldsymbol{h}_{u}^{2}}$, and ${\boldsymbol{h}_{u}^{3}}$). The output of this layer is the graph information vector $\bm{h}_u^4 \in \mathbb{R}^{\lambda_0 \times 1}$ of vehicle $u$, which can be expressed as    
    \begin{equation}
    \label{layer_agg}
      {\boldsymbol{h}_{u}^{4}} = \sigma\left( \left[{\boldsymbol{w}_7}{\boldsymbol{h}_{u}^{1}} \|
      {\boldsymbol{w}_8}{\boldsymbol{h}_{u}^{2}}\|
     {\boldsymbol{w}_9}{\boldsymbol{h}_{u}^{3}}
      \right]\right), 
    \end{equation}
    within which $\boldsymbol{w}_7 \in \mathbb{R}^{\left(\Omega_{0}/3 \right) \times \Omega_{0}}$, $\boldsymbol{w}_8 \in \mathbb{R}^{\left(\Omega_{0}/3 \right) \times \Omega_{0}}$, and $\boldsymbol{w}_9 \in \mathbb{R}^{\left(\Omega_{0}/3 \right) \times \Omega_{0}}$ are the weight parameters for the output of hidden layers \uppercase\expandafter{\romannumeral1}-\uppercase\expandafter{\romannumeral3} , respectively. Here, the output ${\boldsymbol{h}_{u}^{4}}$ constitutes the final graph information of vehicle $u$, as it concatenates the graph information of the sampled first, second, and third hop vehicles. We define (\ref{layer_agg}) as a layer aggregation function that is used to combine the output of the node aggregation functions in (\ref{con1}), (\ref{agg2-2}), and (\ref{agg3-2}) for vehicle $u$. We can also consider other types of layer aggregation functions \cite{Zhao2021Search} as shown in Fig. \ref{8-agg}. 
    
    \item \emph{Hidden Layer \uppercase\expandafter{\romannumeral5}-\uppercase\expandafter{\romannumeral7}}: 
    This layer consists of three cascaded FCLs. It finds the relationship between the graph information vector $\boldsymbol{h}_u^4$ and the probability distribution of on-duty service provide vehicle $u$ in the corresponding sensing, communication, or both service modes. 

    \item \emph{Output}: The output is the probability distribution of vehicle $u$ serving $M+N$ service request vehicles in the corresponding sensing or communication mode and it is represneted by $\bm{y}_{u}=\left[y_{u}^{1}, \cdots ,y_{u}^{M+N+1}\right]$. It includes the case that SPV $u$ does not serve any service request vehicles, and hence, we have $\bm{y}_{u} \in \mathbb{R}^{\left(M+N+1 \right) \times 1}$. Based on the probability distribution $\bm{y}_{u}$ of each vehicle $u$, the service mode selection and service request vehicle connection of each SPV is determined by selecting the vehicle with the highest probability.
    Here, the designed GNN model can be applied for a network in which $M$ and $N$ are constant. If $M$ or $N$ in the network changes, we only need to train  hidden layers \uppercase\expandafter{\romannumeral5}-\uppercase\expandafter{\romannumeral7}.  This is because hidden layers \uppercase\expandafter{\romannumeral1}-\uppercase\expandafter{\romannumeral4} are used a fixed number of vehicles to extract the feature of each vehicle.

\end{itemize}

\subsection{Training of the Dynamic GNN based Method}
In this subsection, we first explain the process of finding appropriate aggregation functions for the proposed dynamic GNN. Then, the loss function is defined and the entire training procedure is introduced. To find the appropriate node and layer aggregation functions for different vehicle topologies, we define a trainable vector $\bm{\theta} \in \mathbb{R}^{(KA_1 + A_2) \times 1}$, which represents the probabilities of selecting each node aggregation function and layer aggregation function, within which $A_1$ is the number of the node aggregation functions that the GNN can select, $A_2$ is the number of the layer aggregation functions that the GNN can select, and $K$ is the number of node aggregation functions used in the proposed dynamic GNN based method. From the definition of $\bm{\theta}$, it is actually used to select $K$ node aggregation functions and one layer aggregation function. We then use binary cross entropy (BCE) to capture the difference between the predicted and actual service request vehicle connection results, which is 
    \begin{equation}
    \begin{split}
    \label{loss}    \mathcal{J}\left(\boldsymbol{w},\boldsymbol{\theta} \right)= \sum_{i=1}^{M+N+1} -z^{i}_{u} \log \delta \left({y}^{i}_{u}\left(\boldsymbol{w},\boldsymbol{\theta} \right)\right) \\
    -\left(1-z^{i}_{u}\right) \log \left(1-\delta \left({y}^{i}_{u}\left(\boldsymbol{w},\boldsymbol{\theta} \right)\right)\right),
    \end{split}
    \end{equation}
 within which $\delta\left(\cdot\right)$ is the sigmoid function, $z^{i}_{u}$ is the label of vehicle $u$ in class $i$ which implies that vehicle $u$ is selected to serve service request vehicle $i$, and  $\bm{w}$ is a vector of all GNN parameters. 

Given (\ref{loss}), we then show how to train the proposed GNN model. The proposed dynamic GNN method is trained with an iterative method that consists of two steps: 1) Joint optimization of $\bm{\theta}$ and $\bm{w}$, and 2) optimization of $\bm{w}$ given $\bm{\theta}$. Specifically, the two steps are elaborated as follows:
\subsubsection{Joint optimization of $\bm{\theta}$ and $\bm{w}$} Let $\mathcal{J}_{\text {tra}}\left(\boldsymbol{w},\boldsymbol{\theta} \right)$ and $\mathcal{J}_{\text {val}}\left(\boldsymbol{w},\boldsymbol{\theta} \right)$ denote the training and the validation loss, respectively. Then, the goal of optimizing $\bm{\theta}$ and $\bm{w}$ is expressed as
   \begin{align}
   \mathop{\mbox{min}}_{\bm{\theta}} \quad
& \mathcal{J}_{\text{val}}\left(\bm{w}^*(\bm{\theta}), \bm{\theta}\right) \label{bi-level} \\
   \mbox{s.t.} \quad
& \bm{w}^*(\bm{\theta})=\operatorname{argmin}_{\bm{w}} \mathcal{J}_{\text{tra}}(\bm{w}, \bm{\theta}).\tag{\ref{bi-level}{a}}\label{inner}
   \end{align}
  Due to the difficulty in finding a closed-form solution for (\ref{inner}), we solve (\ref{bi-level}) and (\ref{inner}) in an iterative manner. Firstly, we explain the process of optimizing $\bm{\theta}$ based on validation data. To reduce the computational overhead for obtaining $\nabla_{\boldsymbol{\theta}} \mathcal{J}_{\text{val}}\left(\bm{w}^*(\boldsymbol{\theta}), \boldsymbol{\theta}\right)$ in (\ref{bi-level}), we use a gradient based approximation method \cite{Liu2019DARTS} to approximate $\nabla_{\boldsymbol{\theta}} \mathcal{J}_{\text{val}}\left(\bm{w}^*(\boldsymbol{\theta}), \boldsymbol{\theta}\right)$ by adapting $\bm{w}$ using only a single training step, which is  
   \begin{equation}
   \nabla_{\boldsymbol{\theta}} \mathcal{J}_{\text{val}}\left(\bm{w}^*(\boldsymbol{\theta}), \boldsymbol{\theta}\right) \approx \nabla_{\boldsymbol{\theta}} \mathcal{J}_{\mathrm{val}}\left(\bm{w}-\xi \nabla_{\bm{w}} \mathcal{J}_{\text {tra}}(\bm{w}, \boldsymbol{\theta}), \boldsymbol{\theta}\right),
   \end{equation}
   within which $\xi$ is the learning rate. Then, $\bm{\theta}$ is updated by using a standard gradient descent method, which is expressed by
   \begin{equation}
   \label{theta}
   \boldsymbol{\theta} \leftarrow \boldsymbol{\theta}-\eta \nabla_{\boldsymbol{\theta}} \mathcal{J}_{\mathrm{val}}\left(\bm{w}-\eta\nabla_{\bm{w}} \mathcal{J}_{\text {tra}}(\bm{w}, \boldsymbol{\theta}), \boldsymbol{\theta}\right),
   \end{equation}
   within which $\eta$ is the learning rate. Then, we explain the update process of $\bm{w}$ based on training data. Using a gradient descent method, GNN parameters $\bm{w}$ is updated by
   \begin{equation}
   \label{w}
   \boldsymbol{w} \leftarrow \boldsymbol{w}-\xi \nabla_{\bm{w}} \mathcal{J}_{\text {tra}}(\bm{w}, \boldsymbol{\theta}).
   \end{equation}
   By iteratively updating (\ref{theta}) and (\ref{w}) until convergence, problem (\ref{bi-level}) can be solved and the well-trained $\bm{\theta}^*$ can be obtained.
   
   \subsubsection{Optimization of $\bm{w}$ given $\bm{\theta}^*$} Based on $\bm{\theta}^*$, the node aggregation functions and layer aggregation function for the GNN are determined. Then, the GNN parameters $\bm{w}$ is tuned on the validation data to further improve the performance of the proposed dynamic GNN based method. The update process of $\bm{w}$ is
   \begin{equation}
   \label{w2}
   \boldsymbol{w} \leftarrow \boldsymbol{w}-\mu \nabla_{\bm{w}} \mathcal{J}_{\text {val}}(\bm{w}, \boldsymbol{\theta}^{*}),
   \end{equation} 
    within which $\mu$ is the learning rate. By updating (\ref{w2}) until convergence, the well-trained $\bm{w}^*$ can be obtained.
Finally, based on the well-trained $\bm{w}^*$, the optimal probabilistic service mode selection and service vehicle connection strategy at each SPV, $\boldsymbol{y}_u$,  can be obtained. The training process of the dynamic GNN is summarized in \textbf{Algorithm 1}.

\begin{algorithm}[t]
\caption{Dynamic GNN based Solution for Solving Problem (\ref{eq:litdiff}).}\label{alg:alg1}
\begin{algorithmic}[1] 
\STATE \textbf{Input:} Vehicle features $\bm{f}$ and connection relationships of each vehicle $v \in {\mathcal{V}}$;
\STATE \textbf{Initialize:} $\boldsymbol{\theta}$ and $\boldsymbol{w}$ are initially randomly generated from uniform distribution;
\FOR {$u = 1 \to U$}
\STATE Sample the first hop vehicles ${\cal{S}}^1 \left(u\right)$, second hop vehicles ${{\cal{S}}^2\left(u\right)}$, and third hop vehicles ${{\cal{S}}^3\left(u\right)}$ for SPV $u$;
\STATE  Aggregate the neighboring graph information from the first hop vehicles ${\cal{S}}^1 \left(u\right)$ of SPV $u$ based on (12)-(13);
\STATE Aggregate the neighboring graph information from the second hop vehicles ${\cal{S}}^2 \left(u\right)$ of SPV $u$ based on (14)-(15);
\STATE Aggregate the neighboring graph information from the second hop vehicles ${\cal{S}}^3 \left(u\right)$ of SPV $u$ based on (16)-(17);
\STATE  Combine the output of the node aggregation functions in (13), (15), and (17) for SPV $u$ based on (18);
\STATE Obtain the graph information $\bm{h}_{u}^4$ of vehicle $u$ and use ${\bm{h}_{u}^4}$ to predict the probability distribution ${\boldsymbol{y}_u}$;
\STATE Calculate loss $\mathcal{J}\left(\boldsymbol{w},\boldsymbol{\theta} \right)$ based on (\ref{loss});
\STATE Update $\bm{\theta}$ based on (\ref{theta});
\STATE Update $\bm{w}$ based on (\ref{w});
\ENDFOR
\STATE Obtain the optimal $\bm{\theta^*}$;
\STATE Optimize $\bm{w}$ based on $\bm{\theta^*}$ according to (\ref{w2});
\STATE Determine the probability distribution $\bm{y}_u$ of SPV $u$ providing service for each service request vehicle in the corresponding sensing, communication, or both service modes according to $\bm{\theta^*}$ and $\bm{w}^*$;
\STATE \textbf{Output:} The service mode selection and service request vehicle connection (i.e., $\bm{\alpha}$ and $\bm{\beta}$) for each SPV $u \in {\mathcal{U}}$. 
\end{algorithmic}
\label{alg1}
\end{algorithm}

 \subsection{Implementation of the Dyanamic GNN based Method} 
Next, we analyze the implementation of the dynamic GNN based algorithm. 
Within the considered vehicular network, the proposed dynamic GNN is implemented on a controller, and control the vehicles with  
1) each vehicle reports its GPS data to the central controller such that the central controller can obtain the vehicle network topology, and 2) the central controller sends decision results of service mode selection and service request vehicle connection to each vehicle.
To use the dynamic GNN based algorithm to solve problem (\ref{eq:litdiff}), the central controller must first calculate the molecular absorption path gain $H^{\textrm{B}}_{um}$, free space path gain $H^{\textrm{F}}_{um}$, effective antenna gain $A_{um}^{\textrm{T}}$ of SPV $u$ transmitting data to communication service request vehicle $m$, and effective antenna gain $A_{um}^{\textrm{R}}$ of communication service request vehicle $m$ served by SPV $u$. To calculate the path gains and effective antenna gains, the central controller needs to know the geographical location information of each vehicle and the antenna direction of each pair of vehicles. The central controller can obtain the geographical location information and antenna direction information according to vehicles' periodically reported GPS data. 
Using the obtained geographical location information and antenna direction information of each vehicle, we can construct the vehicle network topology and then formulate a graph to represent the vehicle network topology. Based on this graph representation, a dynamic GNN will be deployed to output the optimal service mode selection and service request vehicle connection strategy of each SPV $u$ (i.e., $\bm{\alpha}$ and $\bm{\beta}$).
After the central controller determining the service mode selection and service request vehicle connection, it sends the decision results to each vehicle. Since the size of GPS data (i.e., latitude, longitude, and direction) and decision results (i.e., service mode selection and service request vehicle connection matrices) are relatively small, we can neglect the communication overhead and delay used for GPS data and decision results transmission.

 \subsection{Complexity Analysis}
The complexity of the dynamic GNN based method depends on training the dynamic GNN model (i.e., optimizing the GNN parameters $\bm{\theta}$ and $\bm{w}$). The computational complexity of training $\bm{\theta}$ depends on the number $A_1$ of available node aggregation functions, the number $A_2$ of available layer aggregation functions, and the number $K$ of required node aggregation functions. Hence, the computational complexity for training $\bm{\theta}$ is ${\mathcal{O}}\left(A_2 A_1^K\right)$. The computational complexity of training $\bm{w}$ depends on the width, depth, and number of parameters in the GNN model. These GNN parameters are determined by the selected node and layer aggregation functions. For example, if we use the node and layer aggregation functions in Fig.~\ref{nn}, the complexity of training $\bm{w}$ is ${\cal{O}} \left(\left( \Omega_{0} {|\cal{U}|} {\prod_{k=1}^{K} |{\mathcal{S}}^k(u)|} \right)\prod_{i=1}^{I} C_i \right)$, within which $C_i$ is the number of the neurons in layer $i$ and $I$ is the number of hidden layers. Therefore, the computational complexity of iteratively updating $\bm{\theta}$ and $\bm{w}$ is 
\begin{equation}
{\cal{O}} \left(\left(A_2A_1^K\right) \left(\Omega_{0} {|\cal{U}|} {\prod_{k=1}^{K} |{\mathcal{S}}^k(u)|} \right)\prod_{i=1}^{I} C_i \right).
\end{equation}
Notice that the proposed dynamic GNN based model will only trained once during the offline training phase, such that this training process will not introduce additional costs in terms of time, energy, or computing resource during the service.
 Compared to current works \cite{Jeon2020SCALE, Lee2022Intelligent,Lee2021Graph} that need to train the entire neural network models whenever $M$ and $N$ in the network change, our method only needs to retrain three fully connected layers since the hidden layers V-VII are three cascaded fully connected layers. In consequence, our designed method has a lower training complexity compared to the current methods \cite{Jeon2020SCALE, Lee2022Intelligent,Lee2021Graph}. To further enhance the scalability of the designed GNN model, we can define the number of neurons in hidden layer VII as the maximum number of service request vehicles i.e., $M_{\text{max}}+N_{\text{max}}$. In this case,  our designed GNN method can also be applied to the scenarios where the number of service request vehicles is less than $M_{\text{max}}+N_{\text{max}}$. To use our GNN model for these scenarios, we only need to deactivate $(M_{\text{max}}+N_{\text{max}}-M-N)$ neurons  in hidden layer VII.

\section{Simulation Results and Analysis}
\label{sec:4}
\begin{figure}[t]
\centering
\includegraphics[width=0.8\linewidth]{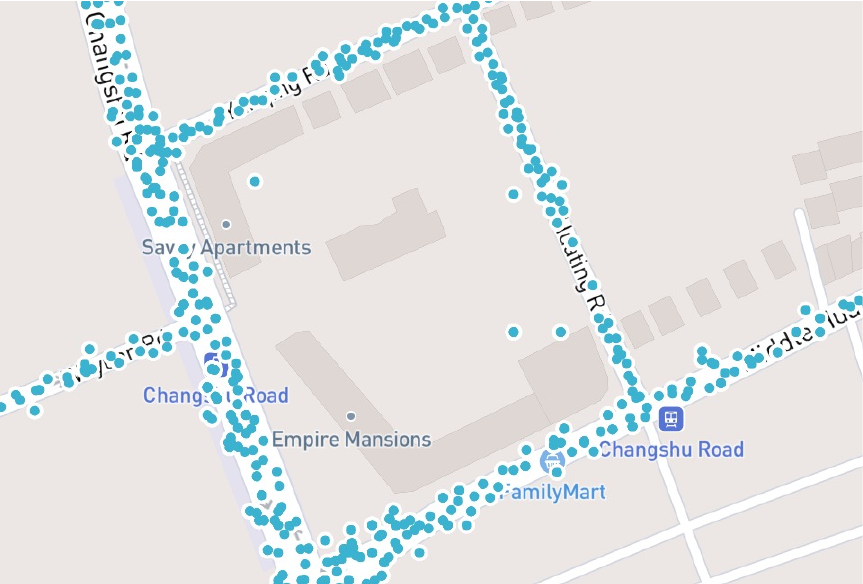}
\caption{Visualization of the GPS data.}
\label{8-1-street}
\end{figure}

For our simulations, we consider an urban area (one street block locates in Shanghai) with size $100$ m $\times$ $100$ m, as shown in Fig. \ref{8-1-street}. Based on the GPS Shanghai Taxi dataset in \cite{Zhao2017Prediction}, we collect and construct $3,500$ vehicle topologies within this region, with the time interval between two vehicle topologies being $30$ seconds. $1,500$ out of these $3,500$ vehicle topologies are considered as the training dataset, $1,000$ of the vehicle topologies are the testing dataset, and the rest $1,000$ vehicle topologies function as the validation dataset.  For each vehicle topology, it contains $U$ data points with $U$ being the number of service provider vehicles (SPVs) in the vehicle topology. Each data point consists of  1) graph information of SPV $u \in \mathcal{U}$, 2) graph information of neighboring vehicles of SPV $u \in \mathcal{U}$, and 3) the actual service request vehicle connection result of SPV $u\in \mathcal{U}$. The other parameters used in simulations are listed in Table \uppercase\expandafter{\romannumeral2} \cite{Shafie2021Coverage}. The results of proposed dynamic GNN based algorithm are compared with the ones of an exhaustive search algorithm (noted as baseline a), a standard GNN algorithm with a fixed neural network model (noted as baseline b), and an optimization solution that directly uses the geographic location information (noted as baseline c). Notice that the exhaustive search algorithm can find the optimal solution for the considered problem. Thus, the comparison between baseline a and the proposed solution can testify the optimality of the results of the proposed solution. Meanwhile, the comparison between baseline b and the proposed solution demonstrates how dynamically choosing different aggregation functions for different vehicle network topologies can improve the performance of the joint sensing and communication services in regards to successfully served vehicles. 
Meanwhile, the comparison between baseline c and the proposed solution can justify the benefits for using GNNs to extract more comprehensive vehicle topological and geographical information.

\begin{table}[!t]
\caption{System Parameters\label{tab:table2}}
\centering
\begin{tabular}{|c|c|c|c|c|c|}
\hline
\!\textbf{Parameters}\! \!\!& \textbf{Value} &\! \textbf{Parameters} \!& \textbf{Value} \\
\hline
$c$ & $3 \times 10^{8}$ m/s & $f$ & $1.05$ THz 
\\
\hline
$P$ & $40$ dBm & $B$ & $5$ GHz \\
\hline
$\varepsilon_{0}$  & $-77$ dBm  & $\tau(f)$ & $0.07512$ m$^{-1}$ \\
\hline 
$\iota$ & $0.1$ & $\varrho, \varsigma$  & $10^\circ$, $10^\circ$ \\
\hline
$\kappa$ & $1$ & $\gamma_{\min}$ & $3$ dB \\
\hline
$Q$ & $20$ MB & $D_{max}$ & $5$ ms \\
\hline 
$\Omega_0$ & $64$ & $p$ & $0.9$ \\
\hline 
$|{\mathcal{S}}^1(u)|$ & $10$ & $|{\mathcal{S}}^2(u)|$ & $10$ \\
\hline 
$|{\mathcal{S}}^3(u)|$ & $10$ & $K$ & $3$ \\
\hline 
$A_1$ & $8$ & $A_2$ & $3$ \\
\hline 
\multicolumn{2}{|c|}{Size of hidden layer \uppercase\expandafter{\romannumeral5}-\uppercase\expandafter{\romannumeral7}} & \multicolumn{2}{|c|}{$32$, $64$, $64$} \\
\hline  
\end{tabular}
\end{table}

\begin{figure}[t]
\centering
\setlength{\belowcaptionskip}{-0.5cm}
\includegraphics[width=0.9\linewidth]{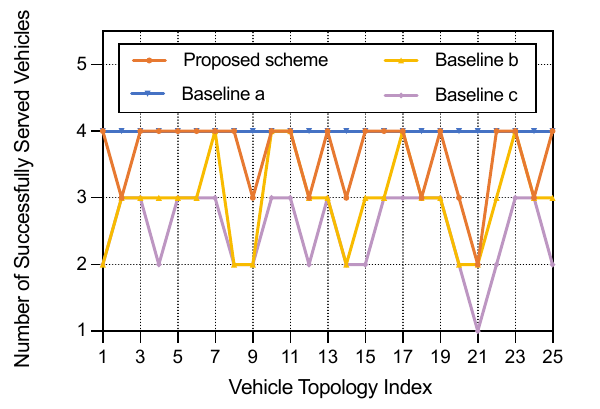}
\caption{The number of successfully served vehicles as the vehicle topology varies ($U=10$, $M=2$, and $N=2$).}
\label{8-success}
\end{figure} 

Fig. \ref{8-success} shows how the number of successfully served service request vehicles changes with the vehicle network topology. From Fig. \ref{8-success}, we see that the proposed dynamic GNN based method improves the number of successfully served vehicles by up to $17\%$ compared to the standard fixed GNN method (i.e. baselines b). This is due to the fact that the proposed method can learn more graph information of vehicle network topologies by choosing appropriate aggregation functions for different vehicle network topologies. Meanwhile, the proposed method also yields a $28\%$ higher number of successfully served vehicles compared to the optimization method in baseline c, as it uses a GNN model to extract both vehicle location and topology information. Fig. \ref{8-success} also shows that the proposed dynamic GNN based method can achieve $91\%$ of the overall optimal number of successfully served vehicles achieved by baseline a, which justifies that the proposed method is capable of quickly adapting to changes on the vehicle topologies.

\begin{figure}[t]
\centering
\setlength{\belowcaptionskip}{-0.5cm}
\includegraphics[width=0.9\linewidth]{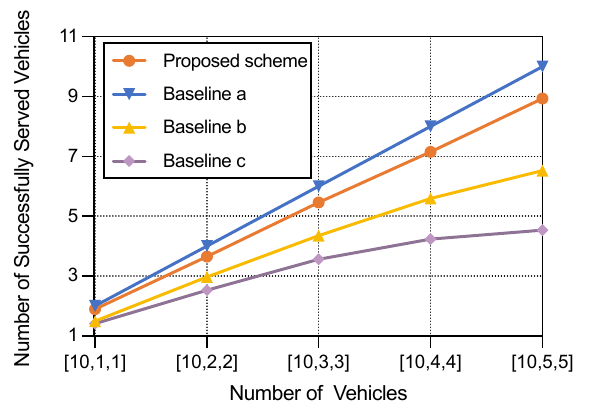}
\caption{The number of successfully served vehicles as the number of vehicles varies ($U=10$, $M$ and $N$ vary from $1$ to $5$).}
\label{8-target}
\end{figure}

Fig. \ref{8-target} shows how the number of successfully served vehicles changes with the number of service request vehicles. In particular, it shows that the number of successfully served vehicles increase with the number of service request vehicles, as the availability of sensing and communication services also increases. Fig. \ref{8-target} shows that, compared to baselines b and c, the proposed method can yield, respectively, up to $19.79\%$ and $32.84\%$ higher number of successful served vehicles. These gains stem from the fact that the adopted dynamic GNN model can extract more vehicle network information than the fixed neural network model, and can inclusively extract the connectivity information of all vehicles. Fig. \ref{8-target} also shows that the difference between the number of successfully served vehicles results from proposed method and one results from baseline a is less than $8.85\%$. This further justifies that the proposed methods can adapt to various vehicle network topologies within which the number of service request vehicles changes.

\begin{figure}[t]
\centering
\setlength{\belowcaptionskip}{-0.5cm}
\includegraphics[width=0.9\linewidth]{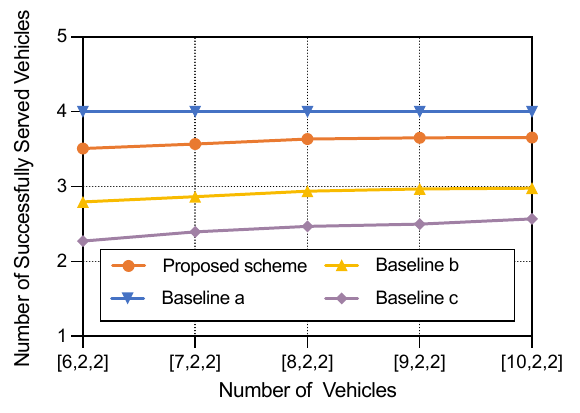}
\caption{The number of successfully served vehicles as the number of SPVs varies ($M=2$, $N=2$, and $U$ varied from $6$ to $10$).}
\label{8-4-spv}
\end{figure}

Fig. \ref{8-4-spv} compares the adaptability of all considered algorithms under different vehicle network topologies with different number of SPVs. In particular, it shows the number of successfully served vehicles increases with the number of SPVs. This is because more SPVs can provide more connection options for service request vehicles, thereby increasing the probability of meeting sensing and communication service requirements. From Fig. \ref{8-4-spv}, we also see that the proposed method yields up to $17.35\%$ and $29.1\%$ gains in terms of the number of successfully served vehicles compared to baselines b and c. The $17.35\%$ gain stems from the fact that the proposed method can capture the dynamics of vehicle network topologies caused by vehicle movements and dynamic wireless channels. The $29.1\%$ gain is because the proposed GNN based method considers topological information of all vehicles to manage the interference among sensing and communication links.

\begin{figure}[t]
\centering
\setlength{\belowcaptionskip}{-0.5cm}
\includegraphics[width=0.9\linewidth]{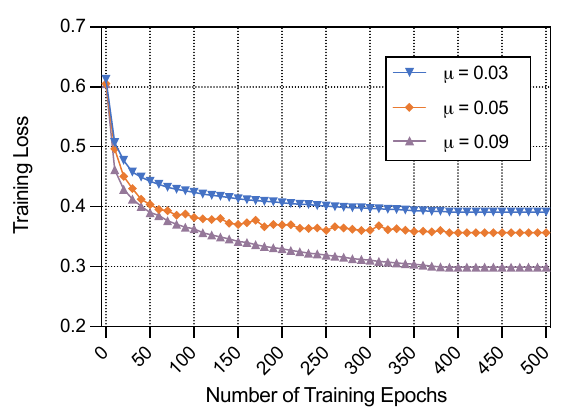}
\caption{The training loss as the number of training epochs varies.}
\label{8-5-loss}
\end{figure}

Fig. \ref{8-5-loss} shows the convergence of the proposed dynamic GNN based method under different learning rates. From Fig. \ref{8-5-loss}, we observe that, as the number of training epochs increases, the training loss of the proposed method decreases first and, then remains stable. This demonstrates that the proposed dynamic GNN based method can reach convergence after the sufficient training epochs (approximately $500$ training epochs). Fig. \ref{8-5-loss} also shows that the proposed method with learning rate being $\mu=0.09$ can reduce training loss by $16.19\%$ and $23.55\%$, compared to the proposed method with learning rate being $\mu=0.05$ and $\mu=0.03$, respectively. This is due to the fact that the proposed dynamic GNN model with learning rate being $\mu=0.09$ can find a better set of weights, and hence, better service mode and service vehicle connection strategy can be determined for each SPV. 

\begin{figure*}[!t]
\centering
\subfloat[GNN model for vehicle network topology with $U=5$, $M=2$, and $N=2$.]{\includegraphics[width=0.9\linewidth]{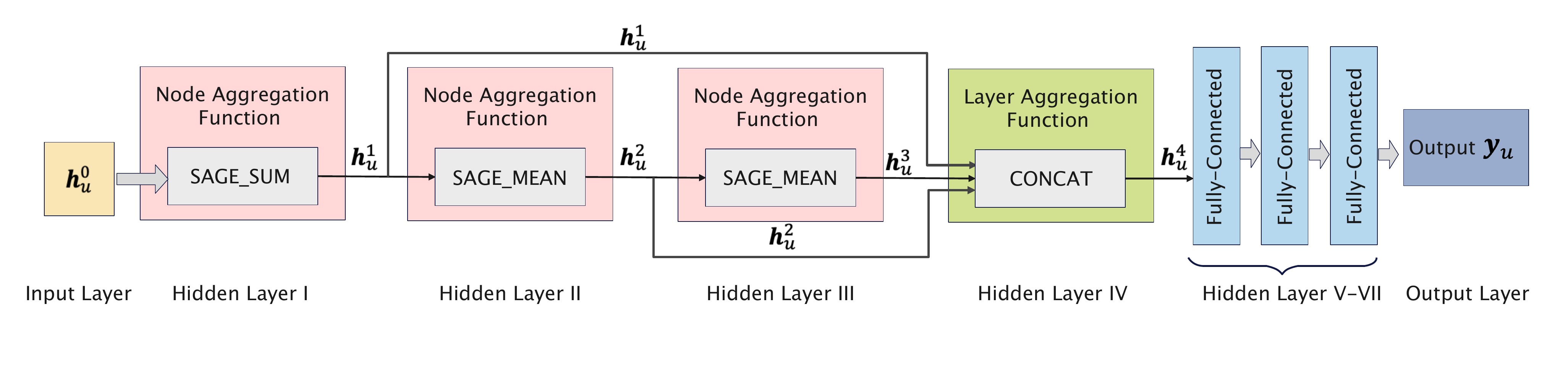}
\label{8-6a}}
\\
\subfloat[GNN model for vehicle network topology with $U=10$, $M=5$, and $N=5$.]
{\includegraphics[width=0.9\linewidth]{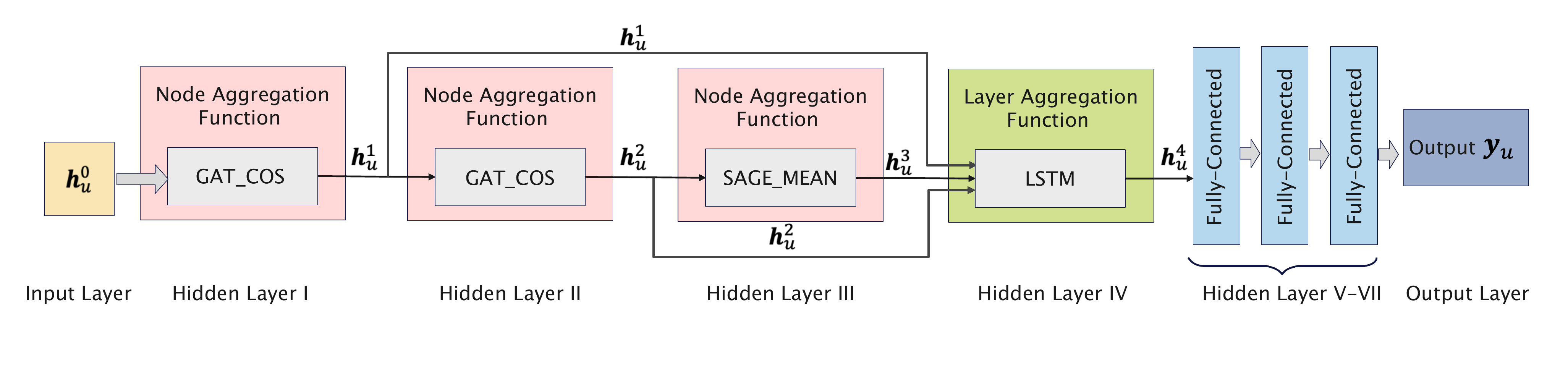}
\label{8-6b}}
\caption{The selected GNN models for different vehicle network topologies.}
\label{8-6-select}
\end{figure*}

Fig. \ref{8-6-select} shows the selected GNN models for different vehicle network topologies. From Fig. \ref{8-6-select}(a), we see that, a simple node aggregation function (e.g., SAGE\_SUM and SAGE\_MEAN) and layer aggregation function (e.g., CONCAT) are selected in a sparse vehicular network with $5$ SPVs, $2$ communication service request vehicles, and $2$ sensing service request vehicles. This is due to the fact that a sparse vehicle network topology only contains limited vehicle topological information. Hence, using complex aggregation functions to extract the topological information of these vehicles may lead to overfitting. From Fig. \ref{8-6-select}(b), we see that, attention mechanism based node aggregation functions (e.g., GAT\_COS) are selected in a dense vehicular network that consists of $10$ SPVs, $5$ communication service request vehicles, and $5$ sensing service request vehicles. This is because attention mechanism based node aggregation functions can measure the importance of different neighborhood vehicles, and hence, accurately extract the graph information of vehicles.

\begin{figure}[t]
\centering
\vspace{-0cm}
\setlength{\belowcaptionskip}{-0.45cm}
\includegraphics[width=0.9\linewidth]{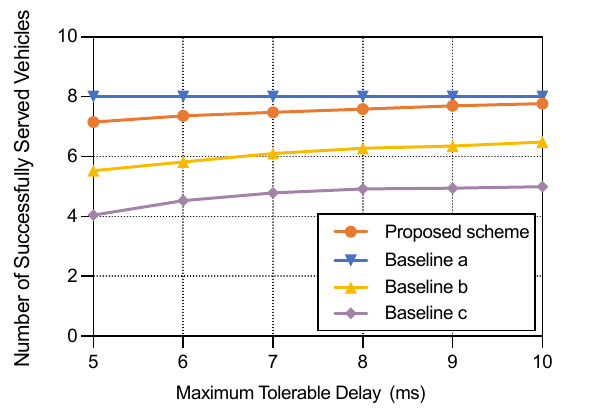}
\caption{The number of successfully served vehicles as the maximum tolerable delay $D_{max}$ varies ($U=10$, $M=4$, and $N=4$).}
\label{8-9-delay}
\end{figure}

Fig. \ref{8-9-delay} shows how the number of successfully served vehicles changes as the maximum tolerable delay of the communication service varies. From Fig. \ref{8-9-delay}, we see that, the number of successfully served vehicles increases as the maximum tolerable delay of the communication service increases. This is due to the fact that, as the maximum tolerable delay increases, one SPV can provide more services that meet the requirement of communication service to service request vehicles. Fig. \ref{8-9-delay} shows that the proposed method can achieve up to $17.64\%$ and $35.03\%$ gains in terms of the number of successful served vehicles compared to baselines b and c, respectively. The $17.64\%$ gain stems from the fact that the proposed method uses different aggregation functions to extract vehicle network information for different vehicle network topologies, and hence, more vehicle network information can be extracted. The $35.03\%$ gain is because the proposed method optimizes the service mode selection and service request vehicle connection by taking the topology related features into consideration.
Fig. \ref{8-9-delay} also shows that the proposed dynamic GNN based method can reach the similar performance as baseline a, which verifies that the proposed method can find a near optimal solution using the dynamic GNN model.

\begin{figure}[t]
\centering
\vspace{-0cm}
\setlength{\belowcaptionskip}{-0.45cm}
\includegraphics[width=0.9\linewidth]{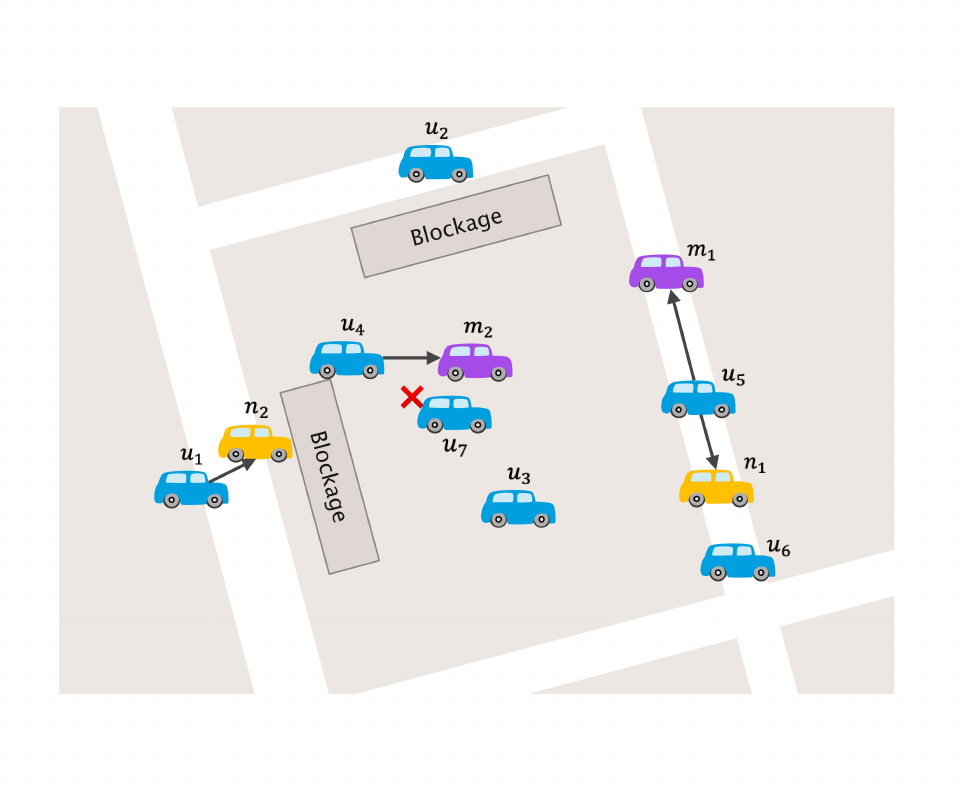}
\caption{Visualization of service mode selection and service request vehicle connection obtained by the proposed dynamic GNN based method.}
\label{8-asso}
\end{figure}


Fig. \ref{8-asso} shows an example of using our proposed method to determine the mode of SPVs and service request vehicle connection. In this figure, SPV $u_7$ is in deactivate state and other SPVs are in active state. From Fig. \ref{8-asso}, we see that the proposed dynamic GNN based method is not only based on vehicle geographical location but also based on sensing and communication interference to determine the service request vehicle connection. For example, in Fig. \ref{8-asso}, sensing service request vehicle $n_1$ does not select SPV $u_6$ to establish a sensing link, although SPV $u_6$ is the closest SPV to sensing service request vehicle $n_1$. The reason is that the establishment of sensing link between vehicle $n_1$ and vehicle $u_6$ will cause interference to the communication link between vehicle $u_5$ and vehicle $m_1$. 
From this figure, we can also see that if the communication service request vehicle and sensing service request vehicle are located in different directions of the SPV, the SPV prefers to provide both communication and sensing services to them. For example, SPV $u_5$ provides communication service to vehicles $m_1$ and sensing service to vehicle $n_1$ simultaneously.

\begin{figure}[t]
\centering
\includegraphics[width=0.9\linewidth]{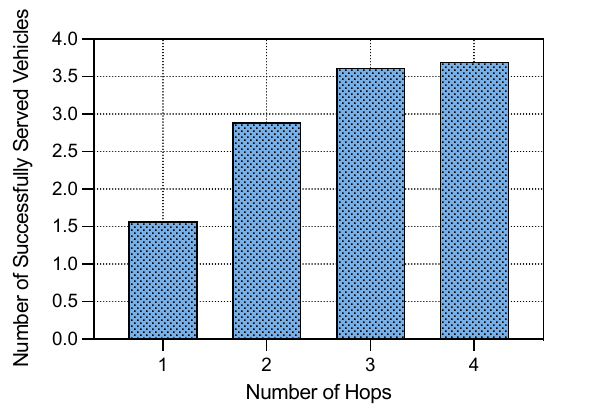} 
\caption{ The number of successfully served vehicles as the number of hops of neighboring vehicles varies ($U=10$, $M=2$, and $N=2$).}
\label{hop}
\end{figure} 
Fig. \ref{hop} shows how the number of successfully served vehicles changes as the number of hops of neighboring vehicles varies. From this figure, we see that, the number of successfully served vehicles improves 18.25\% when the number of hops of neighboring vehicles increases from $2$ to $3$, but only increases 2\% when the number of hops of neighboring vehicles increases from $3$ to $4$. This is due to the fact that the information of three hop vehicles can represent the key geographical information and topological information of a vehicle. 

\begin{figure}[t]
\centering
\includegraphics[width=0.9\linewidth]{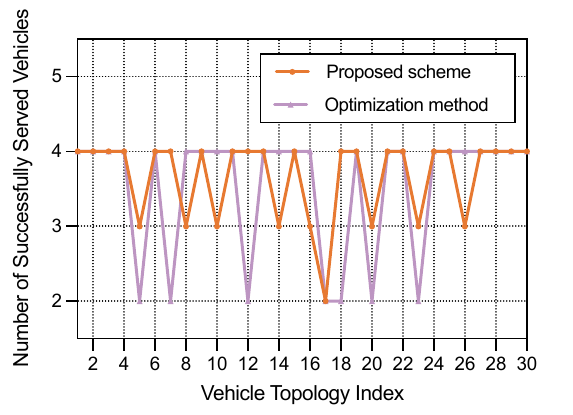}
\caption{The number of successfully served vehicles changes with the vehicle network topology ($U=10$, $M=2$, and $N=2$).}
\label{opt}
\end{figure}
Fig. \ref{opt} shows how the number of successfully served vehicles changes as the vehicle network topology varies. In this figure, we compare the proposed method with a traditional optimization method to solve problem (\ref{eq:litdiff}). In particular, to enable the use of an optimization method to solve problem (\ref{eq:litdiff}), we introduce two auxiliary variables (i.e., $\bm{x}$ and $\bm{\varphi}$) and reformulate the objective function in problem (\ref{eq:litdiff}) as $\mathop{\mbox{max}}_{\bm{\alpha},\bm{\beta},\bm{x},\bm{\varphi}} \quad
\sum_{u\in  \mathcal{U}}\sum_{m\in  \mathcal{M}}x_{um}+\sum_{u\in  \mathcal{U}}\sum_{n\in  \mathcal{N}}\varphi_{un}$, within which $\bm{x}=\left[\bm{x}_{1}, \cdots, \bm{x}_{M}\right]$ with $\bm{x}_{m}=\left[{x}_{1m}, \cdots, {x}_{Um}\right]$ and $\bm{\varphi}=\left[\bm{\varphi}_{1}, \cdots, \bm{\varphi}_{N}\right]$ with $\bm{\varphi}_{n}=\left[{\varphi}_{1n}, \cdots, {\varphi}_{Un}\right]$. $x_{um} = 1$ represents that  communication service request vehicle $m$ is successfully served by SPV $u$; otherwise, $x_{um} = 0$. Similarly, $\varphi_{un} = 1$ represents that sensing service request vehicle $n$ is successfully served by SPV $u$; otherwise, $\varphi_{un} = 0$. 
The constraints in (\ref{eq:11a}) are rewritten as $x_{um} E_{um}^{\textrm{C}}(\bm{\alpha},\bm{\beta}) \geq \alpha_{um} \frac{Q_m}{D_{max}},\forall m \in \mathcal{M},\forall u \in \mathcal{U}$ and $\varphi_{un} \mathrm{\lambda}_{un}^{\textrm{S}}(\bm{\alpha},\bm{\beta}) \geq \beta _{un}\mathrm{\lambda}_{min}, \forall n \in {\mathcal{N}},\forall u \in \mathcal{U}$.
Then, the problem (\ref{eq:litdiff}) can be solved by Matlab Optimization Toolbox~\cite{MatlabOT}. Fig.~\ref{opt} shows that the proposed scheme can improve the number of successfully served vehicles by up to 3.33\% compared to an optimization method. This is because the proposed scheme uses GNNs to capture the geographical location information and topological information of each vehicle, thus finding better service mode selection and service request vehicle connection strategy.

\begin{figure}[t]
\centering
\includegraphics[width=0.9\linewidth]{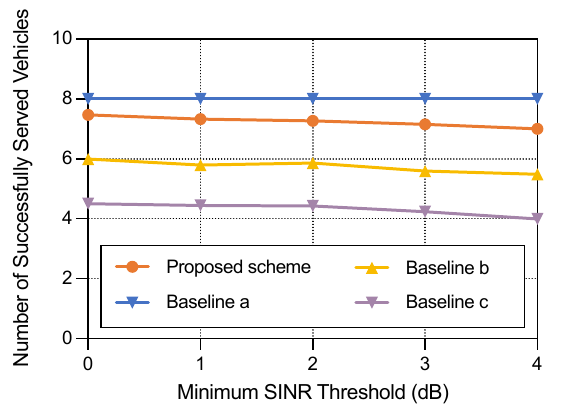}
\caption{ The number of successfully served vehicles as the minimum SINR threshold $\lambda_{min}$ varies ($U=10$, $M=4$, and $N=4$).}
\label{SINR}
\end{figure}
Fig. \ref{SINR} shows how the number of successfully served vehicles changes as the minimum SINR threshold of the sensing service varies. From Fig. \ref{SINR}, we see that, as the minimum SINR threshold of the sensing service increases, the number of successfully served vehicles decreases since the number of service request vehicles that meet the requirement of sensing service decreases. Fig. \ref{SINR} shows that the proposed method can achieve up to 18.71\% and 36.55\% gains in terms of the number of successful served vehicles compared to baselines b and c, respectively. The 18.71\% gain stems from the fact that the proposed method can dynamically select different nodes and layer aggregation functions for different vehicle network topologies, thus improving the performance of the joint sensing and communication services. The 36.55\% gain stems from the fact the proposed method uses a dynamic GNN to extract more comprehensive vehicle topological and geographical information.

\begin{figure*} [t!]
	\centering
    \subfloat[Epoch $0$ ($U=10$, $M=1$, and $N=1$)]{
		\includegraphics[width=0.3\linewidth]{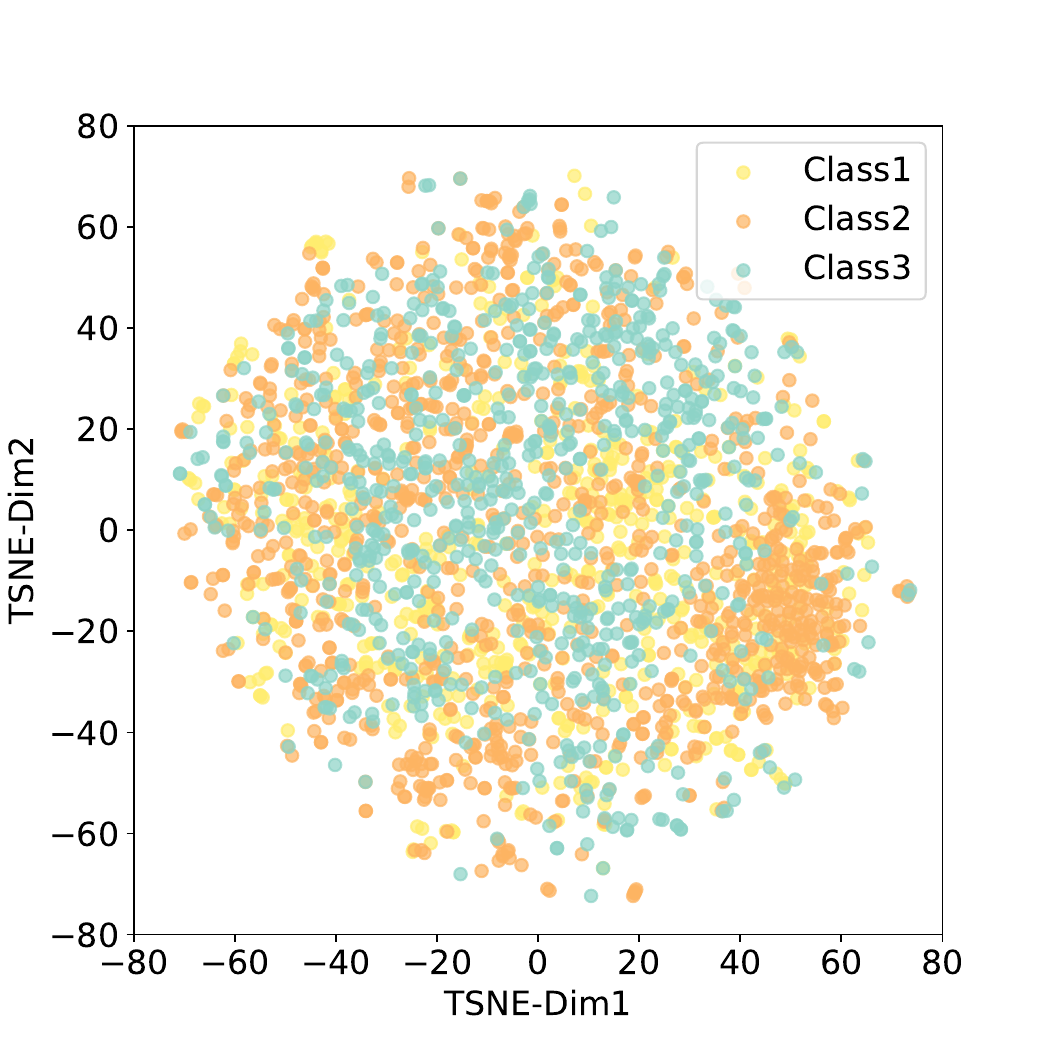}\label{fig:a}}
	\subfloat[Epoch $300$ ($U=10$, $M=1$, and $N=1$)]{
		\includegraphics[width=0.3\linewidth]{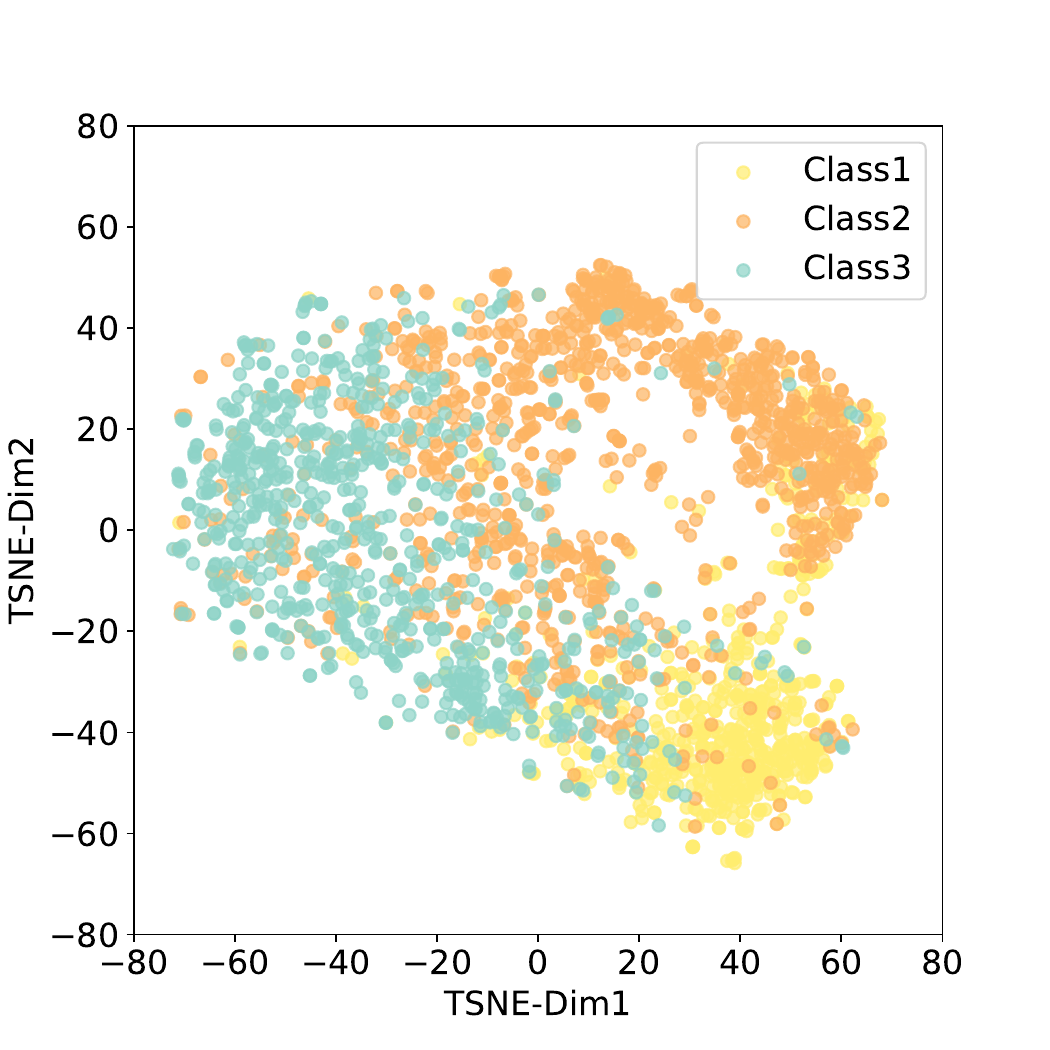}\label{fig:b}}
	\subfloat[Epoch $500$ ($U=10$, $M=1$, and $N=1$)]{
		\includegraphics[width=0.3\linewidth]{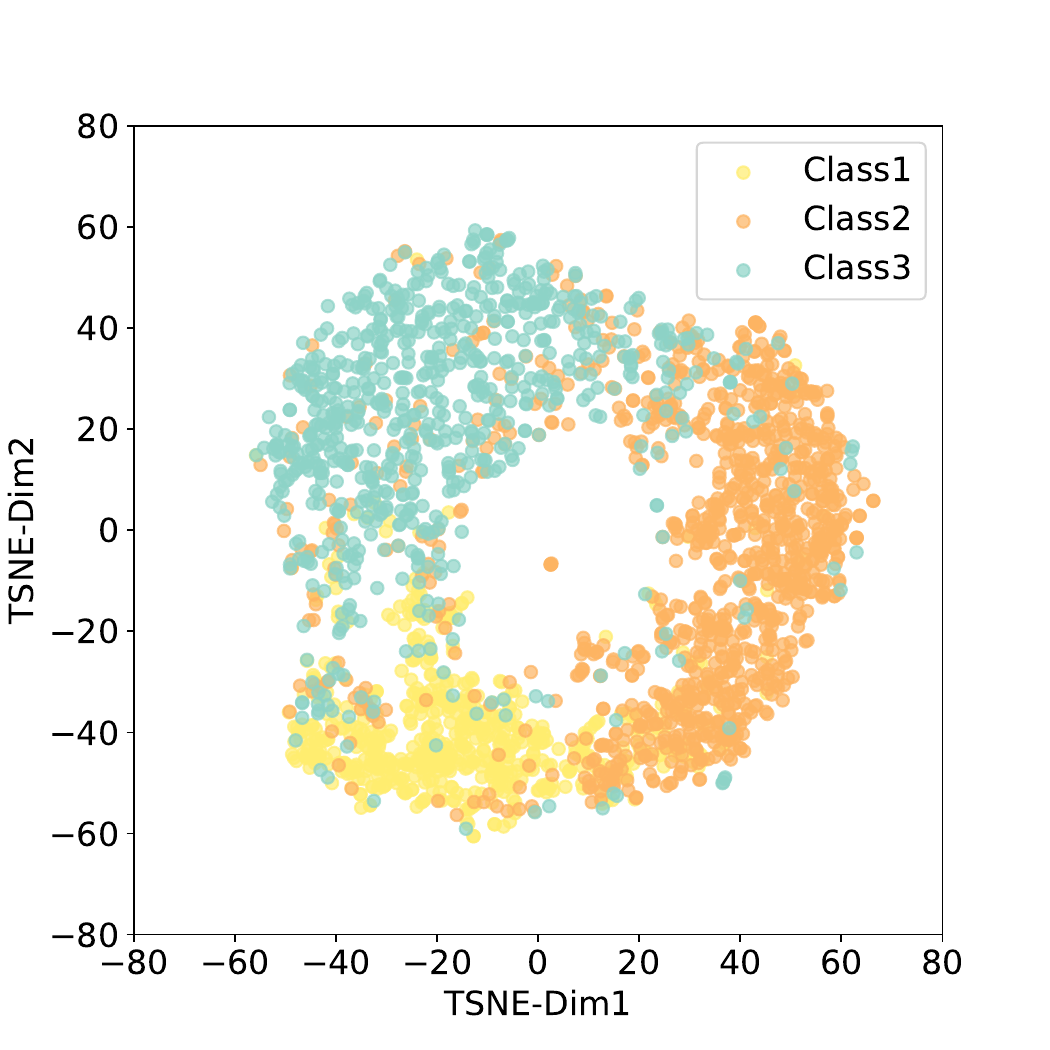}\label{fig:c}}
    \\
    \subfloat[Epoch $0$ ($U=10$, $M=2$, and $N=2$)]{
		\includegraphics[width=0.3\linewidth]{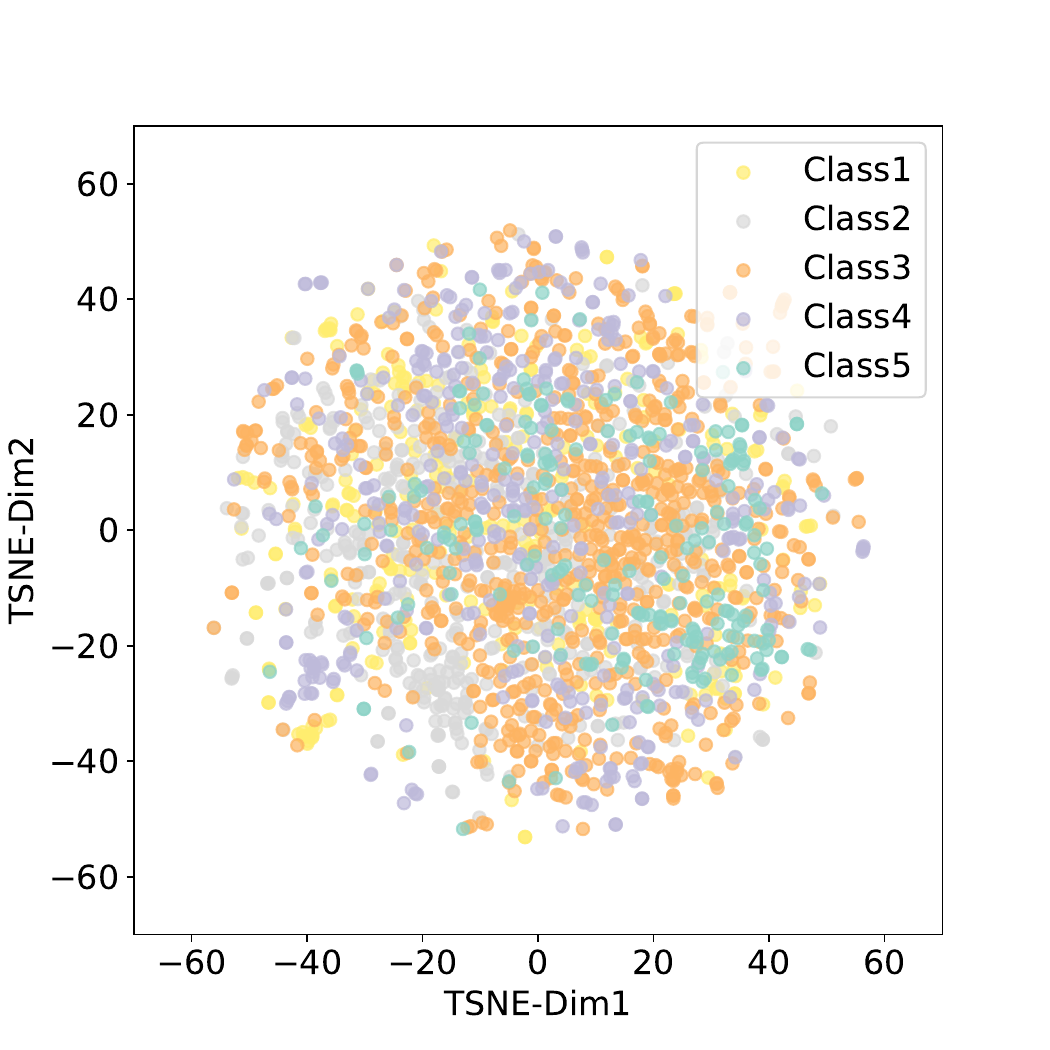}\label{fig:d}}
	\subfloat[Epoch $300$ ($U=10$, $M=2$, and $N=2$)]{
		\includegraphics[width=0.3\linewidth]{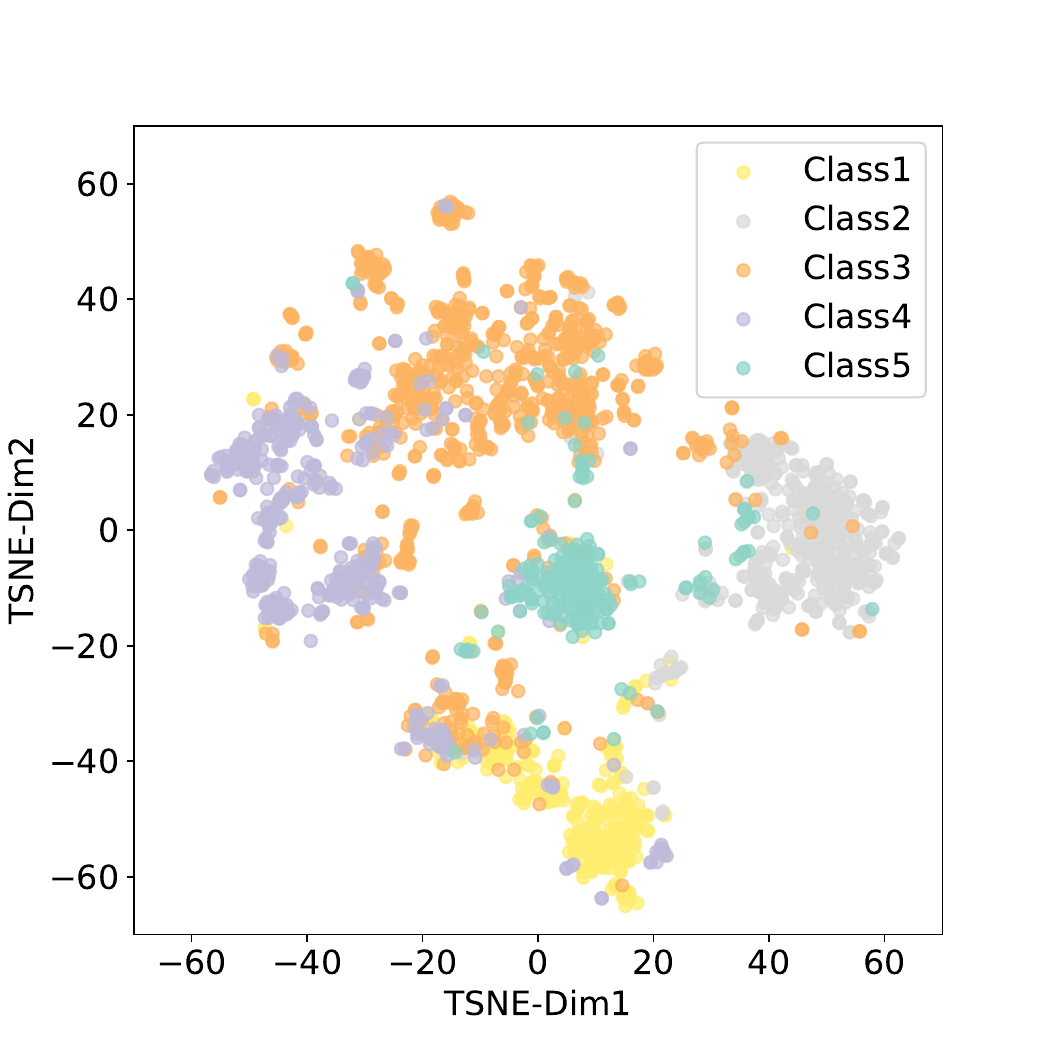}\label{fig:e}}
	\subfloat[Epoch $500$ ($U=10$, $M=2$, and $N=2$)]{
		\includegraphics[width=0.3\linewidth]{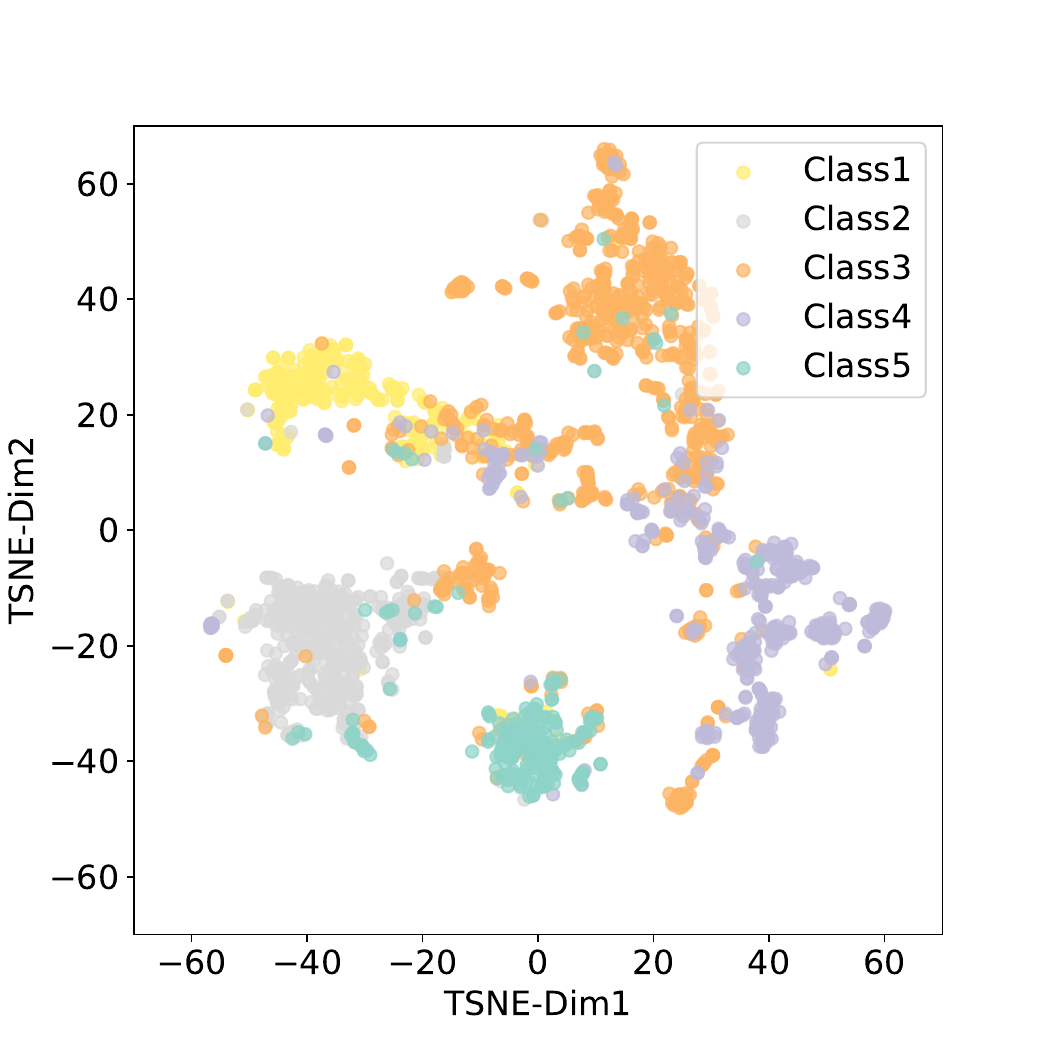}\label{fig:f}}
    \\
	\subfloat[Epoch $0$ ($U=10$, $M=3$, and $N=3$)]{
		\includegraphics[width=0.3\linewidth]{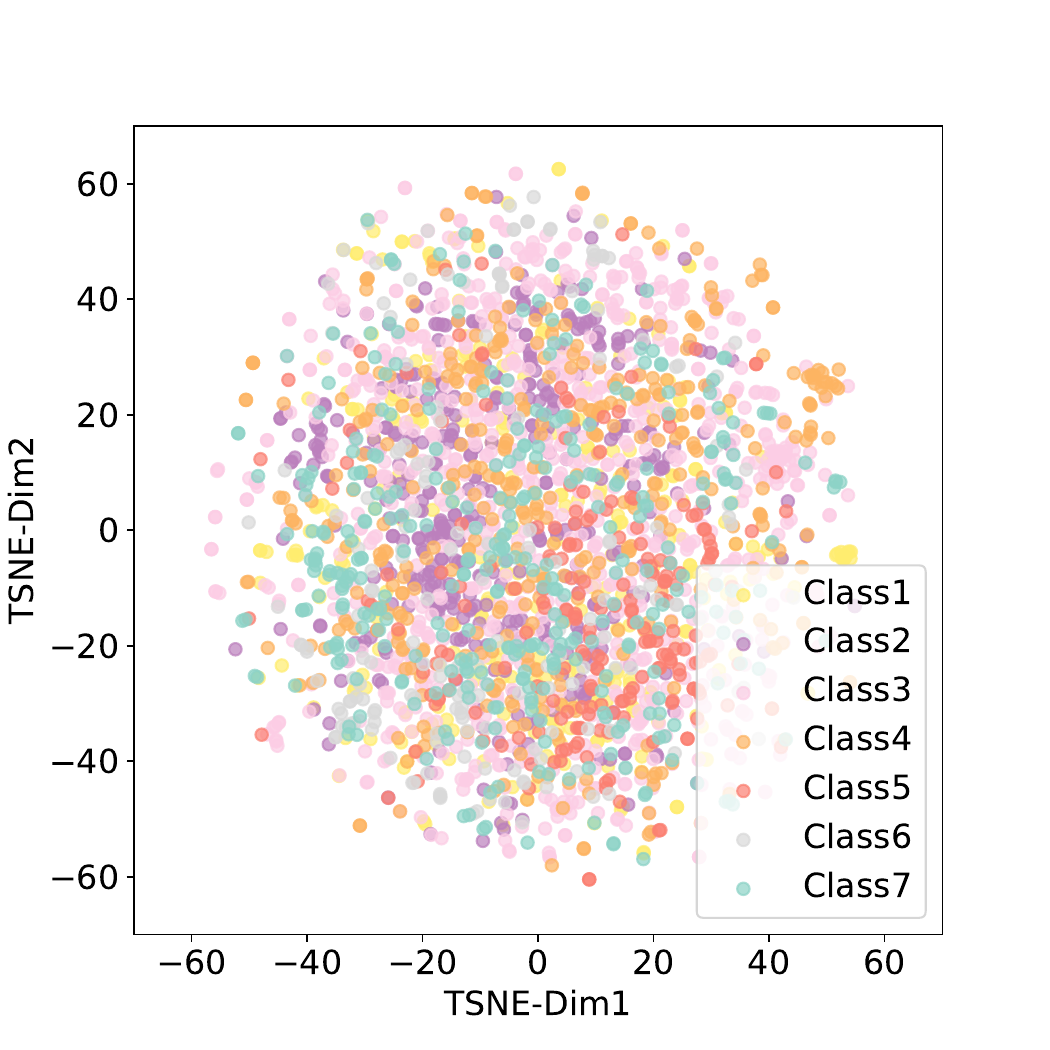}\label{fig:g}}
	\subfloat[Epoch $300$ ($U=10$, $M=3$, and $N=3$)]{
		\includegraphics[width=0.3\linewidth]{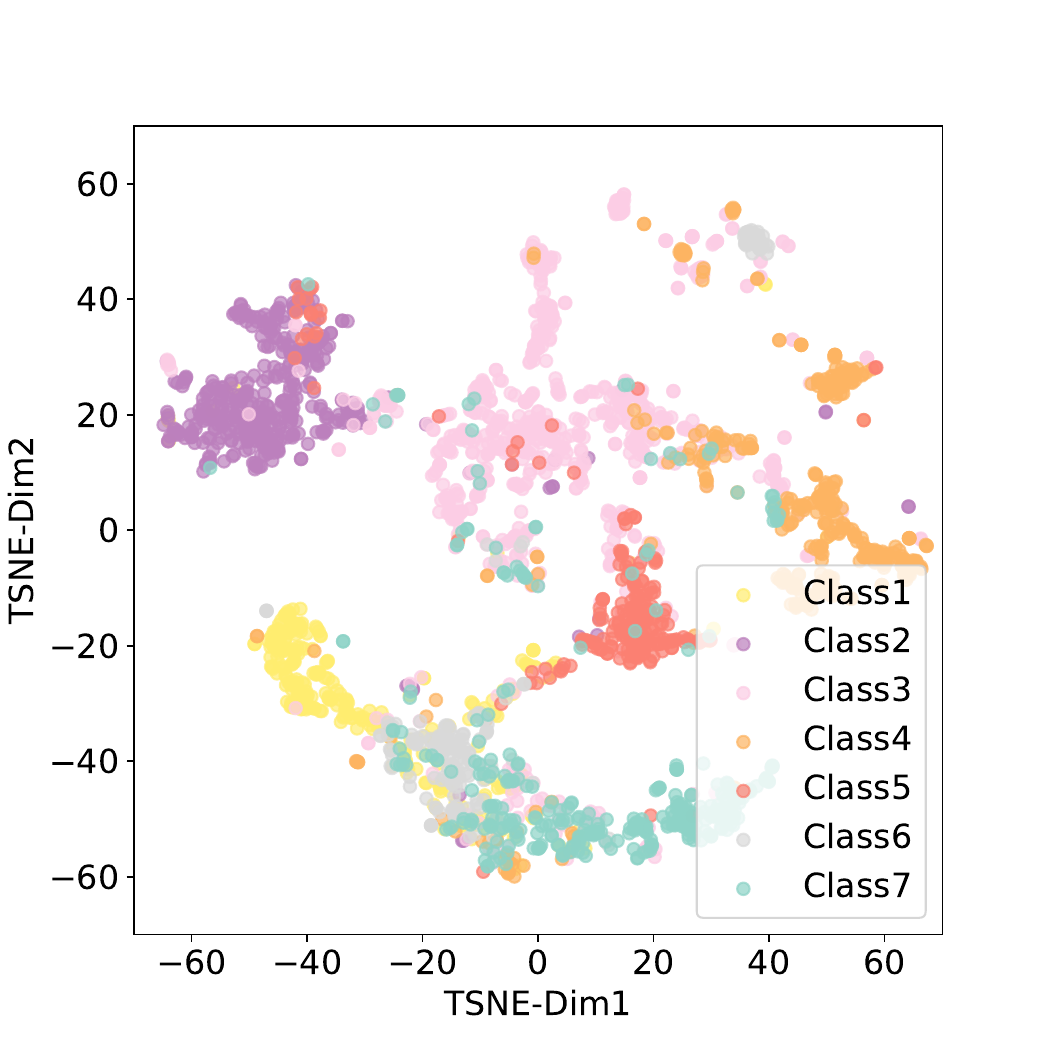}\label{fig:h}}
	\subfloat[Epoch $500$ ($U=10$, $M=3$, and $N=3$)]{
		\includegraphics[width=0.3\linewidth]{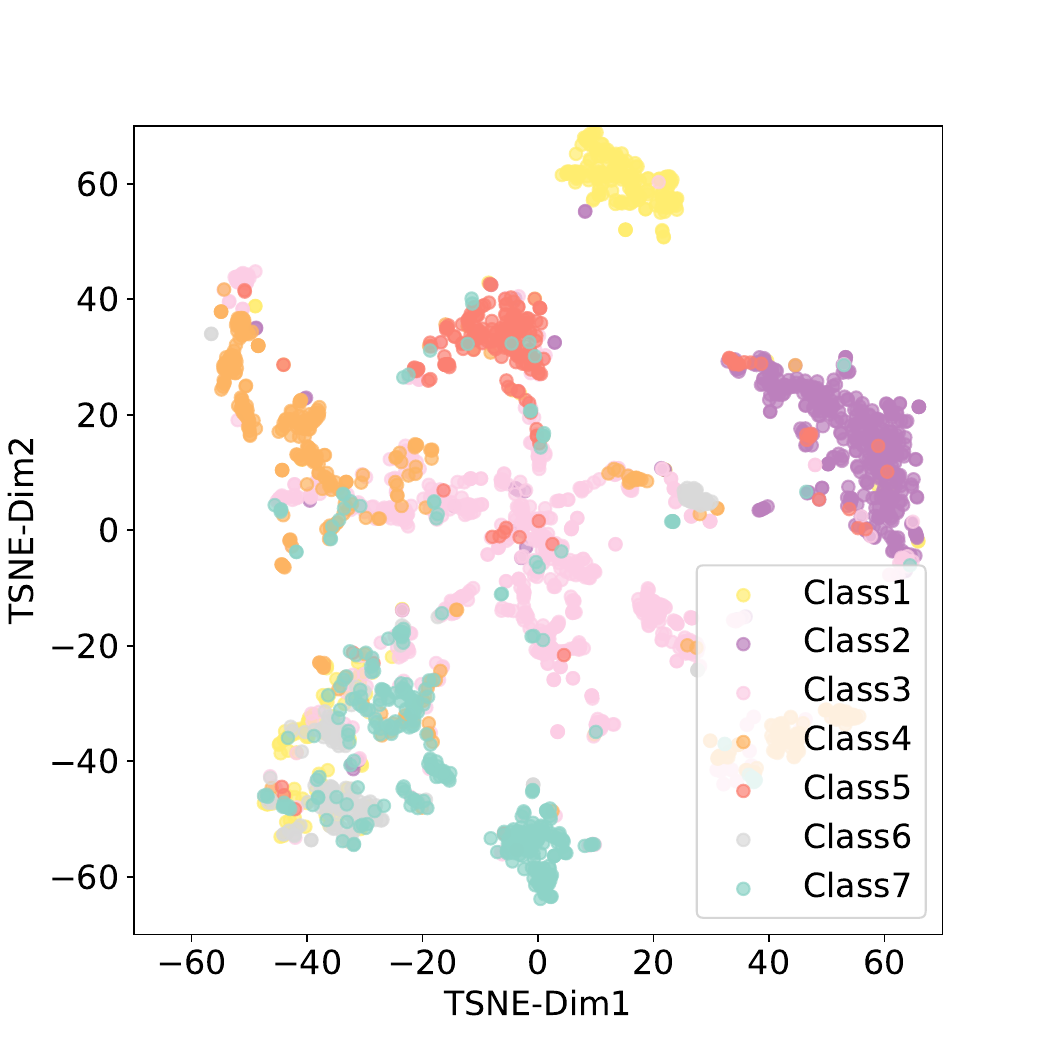}\label{fig:i}}
	\caption{TSNE visualization of graph information vectors for different vehicle network topologies.}
	\label{8-8-emb} 
\end{figure*}

Fig. \ref{8-8-emb} is a visualization of using the proposed GNN model for extracting graph information vectors for different vehicle network topologies. In this figure, the high-dimensional graph information vector ${\boldsymbol{h}_{u}^{4}}$ of each SPV $u$ is mapped into a two-dimensional space by TSNE visualization method \cite{Andrew2023Unexplainable}. The points represent the graph information vectors of SPVs and the colors of points represents the classes of SPVs. In Fig. \ref{8-8-emb}, the points with high similarity are close to each other. For example, as shown in Fig. \ref{8-8-emb}(c), the points of the same class are clustered. This is because the SPVs that trend to provide sensing or communication service have the similar geographical and topological features, and hence, the extracted graph information vectors of them are similar. From Fig. \ref{8-8-emb}(a)-\ref{8-8-emb}(c), we see that, as the the number of training epochs increases, the similarity relationship among different points is more clear. This is due to the fact that the well-trained GNN model can accurately extract the graph information of each SPVs, and learn the relationship between graph information vector and classification categories. From Fig. \ref{8-8-emb}(c), Fig. \ref{8-8-emb}(f), and Fig. \ref{8-8-emb}(i), we also see that, the extracted graph information vector can accurately correspond to the class it belongs to for three different vehicle network topologies. This is because the proposed method uses a dynamic GNN model that can select appropriate aggregation functions for different vehicle network topologies, thus obtaining accurate graph information at different vehicle network topologies.
\section{Conclusion}
\label{sec:5}

In this paper, we have studied the problem of THz enabled joint communication and sensing in vehicular networks. Using this network, the vehicles’ dynamic needs are served by the SPVs, which have both sensing and communication service modes. The design problem is to determine the service mode of each SPV and select the service request vehicles served by each SPV, so as to maximize the number of successfully served vehicles. We have cast this problem into an optimization setting which captures the multi-service mode, service vehicle connection blockage, THz channel particularities, and vehicle network topology dynamics. To solve this problem, we have designed a dynamic GNN based method which selects appropriate graph information aggregation functions for different vehicle network topologies, and thus extracts more vehicle network information, such that it improves service coverage at each SPV. Simulation results verified that the proposed dynamic GNN based method can achieve significant gains in terms of successful services, and adaptability, compared to the standard GNN based solution.

\bibliographystyle{IEEEtran}
\def\baselinestretch{0.92}
\bibliography{Myref}

\vfill

\end{document}